\documentclass{article}
\pdfoutput=1
\usepackage{arxiv}
\usepackage[utf8]{inputenc} 
\usepackage[T1]{fontenc}    
\usepackage[colorlinks,bookmarksopen,bookmarksnumbered,citecolor=red,urlcolor=red]{hyperref}
\usepackage{url}            
\usepackage{booktabs}      
\usepackage{amsfonts}     
\usepackage{nicefrac}     
\usepackage{microtype}     
\usepackage[sort&compress,numbers]{natbib}
\usepackage{doi}
\usepackage{fancyhdr}
\usepackage{graphicx,epstopdf}
\usepackage{amsmath}
\usepackage{amssymb}
\usepackage{multirow}
\usepackage{caption}
\usepackage{subcaption}
\usepackage{xcolor, soul}
\usepackage{adjustbox}
\usepackage[para,online,flushleft]{threeparttable}
\usepackage{enumitem}

\title{A Pole-Based Approach to Interpret Electromechanical Impedance Measurements in Structural Health Monitoring}

\author{\href{https://orcid.org/0000-0002-4694-3237}{\includegraphics[scale=0.06]{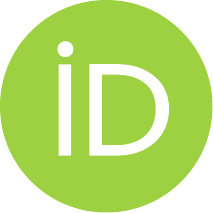}\hspace{1mm}Sourabh Sangle}\thanks{Corresponding author} \\
	J. Mike Walker ’66 Department of Mechanical Engineering\\
        Texas A\&M University\\
        College Station, TX 77843\\
	\texttt{sangle96@tamu.edu} \\
	\and
	\href{https://orcid.org/0000-0001-7864-551X}{\includegraphics[scale=0.06]{orcid.pdf}\hspace{1mm}Sa'ed Alajlouni} \\
	Department of Mechatronics Engineering\\
        Faculty of Engineering\\
        The Hashemite University\\
        Zarqa 13133, Jordan\\
	\texttt{saed@hu.edu.jo} \\
	\and
        \href{https://orcid.org/0000-0002-5443-8484}
    {\includegraphics[scale=0.06]{orcid.pdf}\hspace{1mm}Pablo A. Tarazaga}\\
	J. Mike Walker ’66 Department of Mechanical Engineering\\
        Texas A\&M University\\
        College Station, TX 77843\\
	\texttt{ptarazaga@tamu.edu} \\	
}

\renewcommand{\shorttitle}{A Pole-Based Approach to Interpret EMI in SHM}

\hypersetup{
pdftitle={A Pole-Based Approach to Interpret EMI in SHM},
pdfauthor={Sourabh Sangle},
pdfkeywords={Structural Health Monitoring, Rational Function Approximation, damage detection and diagnosis, Vibrations, Modal parameters},
}

\begin{document}

\maketitle
\begin{abstract}
	Over several decades, electromechanical impedance (EMI) measurements have been employed as a basis for structural health monitoring and damage detection. Traditionally, Root-mean-squared-deviation (RMSD) and Cross-correlation (XCORR) based metrics have been used to interpret EMI measurements for damage assessment. These tools, although helpful and widely used, were not designed with the idea to assess changes in EMI to underlying physical changes incurred by damage. The authors propose leveraging vector fitting (VF), a rational function approximation technique, to estimate the poles of the underlying system, and consequently, the modal parameters which have a physical connection to the underlying model of a system. Shifts in natural frequencies, as an effect of changes in the pole location, can be attributed to changes in a structure undergoing damage. With VF, tracking changes between measurements of damaged and pristine structures is physically more intuitive unlike when using traditional metrics, making it ideal for informed post-processing. Alternative methods to VF exist in the literature (e.g., Least Square Complex Frequency-domain (LSCF) estimation, adaptive Antoulas--Anderson (AAA), Rational Krylov Fitting (RKFIT)). The authors demonstrate that VF is better suited for EMI-based structural health monitoring for the following reasons: 1. VF is more accurate at high frequency, 2. VF estimates complex conjugate stable pole pairs, close to the actual poles of the system, and 3. VF can capture critical information missed by other approaches and present it in a condensed form. Thus, using the selected technique for interpreting high-frequency EMI measurements for structural health monitoring is proposed. A set of representative case studies is presented to show the benefits of VF for damage detection and diagnosis.
\end{abstract}


\section{Introduction}\label{sec:intro}

\quad Industries constantly strive to improve their production either by adopting new practices or improving their current practices. Along with the widely adopted traditional manufacturing techniques, industries are gradually moving towards additive manufacturing (AM) techniques \citep{general_electric_fit_2014,goehrke_rolls-royce_2015,boen_future_2015}. With the expanding production capabilities, the quality control aspect of manufacturing needs to progress at a similar pace. Minimizing damage occurrences or early detection of defects can help these industries lower production time, energy, and thus, the overall production cost. Structural health monitoring (SHM), as the name suggests, deals with monitoring and evaluating the structural health of a part under investigation. For SHM, the state of a structure under investigation is compared against a damage-free baseline state.
Assessment of damage requires comparing two system states - pristine and damaged \citep{worden_fundamental_2007}. With this comparison in mind, any sudden or significant deviations between the two states can be attributed to damage incurred in the part under test. SHM helps in asserting a product's integrity, reliability, and safety. Undetected or incorrectly detected defects may lead to poor quality production, early repairs, or replacements; which in the worst case can lead to the loss of life. Improvements in SHM of manufactured parts, be it in-situ (during production) or post-production, can prove instrumental in increasing the quality and durability of these parts. Thus, improving the current monitoring practices is a path forward toward boosting manufacturing. There exist numerous techniques at one's disposal to perform health monitoring; Radiography Testing (RT), Ultrasonic Testing (UT), and Computed Tomography (CT) X-ray scanning \citep{kamsu-foguem_knowledge-based_2012}, to name a few. 
For this study, the authors focus on electromechanical impedance (EMI) based SHM, a vibration-based technique that has been in use for several decades. EMI-based SHM has emerged as a promising, highly sensitive, and cost-effective non-destructive solution for real-time damage assessment \citep{park_overview_2003}. 
This technique has been implemented for detecting defects for various applications such as civil structures, manufacturing, aerospace structures, composite structures, wind turbines, and space structures \citep{annamdas_electromechanical_2013, xu_modified_2002, taylor_incipient_2013, pitchford_impedance-based_2007,giurgiutiu_piezoelectric_2002, annamdas_application_2010, zagrai_piezoelectric_2010, peairs_frequency_2007,albakri_impedance-based_2017, sturm_situ_2019, sapidis_flexural_2023, naoum_electro-mechanical_2023,tenney_internal_2017,bilgunde_-situ_2018,haydarlar_structural_2022}.
In the context of damage detection, a \emph{damage metric} is a measure used to quantify differences and establish criteria to qualify changes as damage(s). Damage metrics are usually non-dimensional quantities that are compared against a threshold, to binary-classify a part as either "damaged" or "undamaged". The root-mean-squared deviation (RMSD) and the cross-correlation (XCORR) based metrics are commonly used in EMI-based SHM. Commonly, the EMI-based SHM technique has a go, no-go modus operandi or in simpler terms a binary decision classification. Apart from averaging information across an entire frequency range of interest, researchers have also adopted a practice of dividing information into smaller connected frequency windows over which to calculate the damage metrics. This is used as an attempt to link damage to frequency bands. RMSD and XCORR-based metrics being indifferent to the nature of the changes, do not provide information about changes in the physical state of a part, and consequently do not inherently provide intuition about damage. Based on the EMI measurement of a given specimen, which provides a non-parametric model of the specimen, the authors propose estimating (fitting) a parametric model; namely, a rational transfer function. Then a more informative damage metric is obtained by tracking changes in modal parameters (pole locations) of the fitted model. In this work, an attempt to move towards a more informative metric, in this complex problem of damage detection, is proposed.

\subsection*{Paper organization and contributions}
The remainder of this paper is organized as follows - Section \ref{sec:EMISHM} provides a short overview of the electromechanical impedance-based structural health monitoring technique along with standard damage metrics. Section \ref{sec:Needofhour} further discusses the limitations and challenges of the current metrics and a path forward. Section \ref{sec:Polebasedapproach} introduces the pole-based approach, followed by a brief comparison between the available techniques in the literature. Section \ref{sec:case_studies} demonstrates an implementation of the technique and presents a discussion of the results. Finally, the conclusion and future works are presented in Section \ref{sec:ConclusionFutureWorks}.

The main contributions of this paper are organized as follows:
\begin{itemize}

    \item The authors propose leveraging a rational function approximation technique, Vector Fitting (VF), for a more informed interpretation of the EMI measurements (Section \ref{subsec:polemethod}).
    \item The authors briefly compare the selected rational function approximation technique to an industry-adopted parameter estimation technique, LSCF (Section \ref{subsec:LSCF}), and other alternative rational function approximation techniques -- AAA, RKFIT (Section \ref{subsec:AAA_RK}).
    \item The authors demonstrate the use of the selected rational function approximation using simulated  (Section \ref{subsec:simulated}) and experimental (Section \ref{subsec:experimental}) examples.
 
\end{itemize}


\section{Overview of EMI-based SHM \& metrics}
\label{sec:EMISHM}
\citet{park_overview_2003} introduced a vibration-based damage identification technique with EMI measurements at its foundation. This particular vibration-based technique utilizes piezoceramic materials, in particular, the lead zirconate titanate (PZT) wafers, to function as collated actuator-sensor, to actuate a structure and simultaneously measure its dynamic response \citep{liang_coupled_1994, giurgiutiu_characterization_2000}. The coupled electromechanical behavior of these piezoelectric materials simplifies measuring the mechanical impedance of the structure under investigation by relating it directly to the electrical impedance of the piezoelectric transducer.
The fundamental basis of this vibration-based technique is that any damage in the structure will modify the physical characteristics, i.e. mass, stiffness, and damping; of the structure under investigation, which will then be reflected in its EMI measurements made across the PZT wafer. The wafers are either attached to the surface or embedded in the structure during fabrication.

Using the coupled electromechanical behavior of the piezoelectric materials, the electrical impedance of the piezoelectric wafer, $Z$($\omega$) can be related to the mechanical impedance of the structure $Z_{st}$ \citep{liang_coupled_1994}, as, 
\begin{equation}
     Z(\omega) = \left [i \omega \frac{bl}{h}
     \left(\frac{d^2_{13}}{s^E_{11}}
     \left( \frac{\tan{kl}}{kl} \left(\frac{Z_{pzt}}{Z_{pzt} +   Z_{st}} \right) - 1\right) + \epsilon^\sigma_{33}     \right)\right]^{-1},
     \label{eq:piezecoupling}
\end{equation}
where $\omega$ is the frequency of excitation, $Z_{pzt}= -ibhl\left(s^E_{11} \omega \frac{\tan{kl}}{kl} \right)^{-1}$ is the piezoelectric transducer short-circuit impedance, $k=\omega \sqrt{\rho s^E_{11}}$ is the wave number, $d_{13}$ is the piezoelectric coupling coefficient, $s^{E}_{11}$ is the complex mechanical compliance measured at zero electric field, $\epsilon^{\sigma}_{33}$ is the complex permittivity measured at zero stress. Likewise, $\rho$ is the density of the piezoelectric material, and $b, h$ and $l$ are the piezoelectric patch width, thickness, and length, respectively.

\citet{annamdas_electromechanical_2013} used this technique to monitor the health of metallic and non-metallic material structures, while \citet{xu_modified_2002} extended the application to detect debonding in composite patches. \citet{taylor_incipient_2013} carried out crack detection in wind turbine rotor blades using impedance measurements and \citet{pitchford_impedance-based_2007} performed structural health monitoring of wind turbine blades. \citet{giurgiutiu_piezoelectric_2002} monitored structural damage in ageing aircraft structures while \citet{annamdas_application_2010} reviewed the space structure applications of the technique. \citet{zagrai_piezoelectric_2010} suggested using impedance-based SHM to perform pre-launch diagnostics along with monitoring structures during launch and evaluating on-orbit operations. \citet{peairs_frequency_2007} examined a composite boom used in a space reflector with this piezoelectric-based technique. 
\citet{albakri_impedance-based_2017} demonstrated the effectiveness of this technique for the non-destructive evaluation (NDE) of additively manufactured parts. \citet{strutner_inexpensive_2018} proved that the EMI-based technique can be a viable and cheap alternative to traditional CT X-ray scanning. Moreover, \citet{sturm_situ_2019} were able to use the technique to detect in-situ build defects as a layer-by-layer process. More recently, \citet{sapidis_flexural_2023} demonstrated the use of EMI for flexural damage evaluation in fiber-reinforced concrete beams, while \citet{naoum_electro-mechanical_2023} investigated the potential of EMI measurements for detecting cracks in fiber-reinforced concrete prisms. All these applications, though not fully comprehensive, make a case for the EMI-based technique being a strong contender for NDE and SHM of various structures. 

To evaluate the integrity or health of a structure, there is a need for a baseline measurement of the structure [\citet{worden_fundamental_2007} - Axiom II]. A quantitative \emph{damage metric} is essential to quantify the difference between the EMI measurements of the undamaged (or baseline structure) and the damaged [\citet{worden_fundamental_2007} - Axiom IVa]. The quantification between the variations in the measurements of a damaged structure and its baseline is usually carried out using norm-based metrics. \citet{sun_truss_1995} was one of the first researchers to use a norm-based metric wherein a sum of the squared difference between measurements was calculated. This metric gradually evolved into the current RMSD-based metric, that is frequently used for EMI-based health monitoring. Another cumulative metric, used as an alternative, or at times alongside the RMSD-based metric, is the XCORR-based metric. It originates from signal processing and differs from RMSD as it focuses on the overall shape similarity between two signals (not their scale/amplitude, as the signals can have different scales). 
The formulations for these commonly used damage metrics are, 
\begin{subequations}
    \begin{equation}
        RMSD_k = \sqrt{\frac{\sum^{j}_{i}\left( Z_{k,i} - Z_{Baseline,i} \right)^{2}}{\left(Z_{Baseline,i}\right)^2}}, \hspace{5pt} \text{and}
        \label{eq:RMSD}
     \end{equation}

    \begin{equation}
        XCORR_k = 1 - \Bigg|\frac{\sum^{j}_{i}\left( Z_{k,i} - \bar{Z}_{k}\right)\left( Z_{Baseline,i} - \bar{Z}_{Baseline}\right)} {\sqrt{\sum^{j}_{i}\left( Z_{k,i} - \bar{Z}_{k}\right)^2}{\sqrt{\sum^{j}_{i}\left( Z_{Baseline,i} - \bar{Z}_{Baseline}\right)^2}}} \Bigg|,
        \label{eq:XCorr}
    \end{equation}
\end{subequations} where $Z_{k,i}$ is the EMI measurement for $k^{th}$ structure at $i^{th}$ frequency value, $j$ is the highest value of the chosen frequency range for which the EMI measurements are recorded and $Z_{Baseline}$ is the baseline measurement. 
For an RMSD-based metric, the numeric value is directly proportional to the degree of difference between the two measurements; i.e. greater the numeric value, the higher the difference between the two measurements. For the XCORR-based metric, instead of measuring the absolute difference between two measurements, relative changes in one measurement from the other are established. The closer the XCORR-based metric value to zero, the smaller the difference between the two measurements. 

\section{Challenges and scope of improvement} 
\label{sec:Needofhour}
Traditionally, variations between the baseline measurements and the measurements of the structures under investigation are quantified using the statistical tools mentioned in Section \ref{sec:EMISHM}; RMSD and XCORR. Although used extensively, these techniques have shortcomings that need to be addressed. These metrics adopt an aggregate frequency approach and treat the underlying system as mere data points with an assumption that information is uniformly distributed across the frequency range. There exists the practice of creating smaller windows in the frequency range to try and address this shortcoming. This practice of using pre-set windows does not technically average information across the complete testing frequency range, but only across these pre-set windows. Whilst this attempts to solve the issue of uniform distribution by moving towards more local frequency regions, the information is still uniformly distributed across the chosen window. A defect can have compensatory effects in adjacent frequency windows, e.g. sensitive mode peak shifts. When the information is averaged out across the frequency window, the effect may be nullified, making this practice ineffective in providing sufficient information about the changes in the underlying system (examples shown in the following sections). Thus, without any prior knowledge of the dynamic richness of the system or frequency-specific distribution of information, the practice of creating windows can be unproductive at times.

These metrics are scalar and provide negligible to minimum intuition about the damage. Consider a scenario wherein different types of damages occur in two distinct structures, such that these norm-like damage metrics report the same numeric value. The only inference available here is that some change has occurred and no way to differentiate between the two scenarios. Researchers have also attempted to make modifications to the metrics. The RMSD-based metric used by \citet{pitchford_impedance-based_2007}, was scaled to the baseline measurement and corrected for vertical shift between measurements assuming that it does not indicate any damage. All three sensors in the study were able to detect both introduced controlled damages - increased mass and increased stiffness, as well as inducing actual damage at various locations on the structure of interest. However, for certain locations or configuration, the metric values where either not significant or the threshold for damage qualification had to be tuned. Moreover, the damage metrics were also indifferent to the damage types, and reported values did not provide additional information about the damage type. According to \citet{annamdas_three-dimensional_2007-1,annamdas_three-dimensional_2007}, the interpretation of the metrics is subjective to the material of the structure. A 5\% variation in RMSD-metric value is considered a severe indicator for concrete structure but not so for metallic structures. Another concern with using the current metrics that needs to be addressed is the inherent assumption that all sections of frequency are equally relevant and important. \citet{peairs_frequency_2007} addressed the importance of frequency range selection for EMI-based SHM sensitivity and proposed a rule of thumb to test at frequency ranges with high modal density. \citet{annamdas_application_2010} also pointed out how the RMSD value is largely affected by the frequency range of excitation i.e., reported metric values for different frequency ranges, e.g., 0 - 50 kHz and 0 - 100kHz, are different for the same recorded measurement. Given the frequency range sensitivity of the standard metrics, i.e. the values are affected by the chosen range, the effectiveness of the current EMI-based evaluation technique is susceptible to window size and location. According to \citet{giurgiutiu_piezoelectric_2002}, for specific applications, systematic pre-test investigations are required along with determining the most appropriate damage metric.

For better and more efficient health monitoring, one would want to progress along the different stages of SHM, first discussed by \citet{rytter_vibrational_1993}; from damage detection (Stage 1) to localization of damage (Stage 2), to characterization (Stage 3) and finally a diagnosis (Stage 4) \& prognosis (Stage 5) to avoid such damages in the future. With the current capabilities, one can determine the presence of damage or defects. \citet{tseng_non-parametric_2002} presented an experimental example wherein damage investigations were performed using transducers attached at opposite ends. One of the conclusions presented was that the RMSD-based metric (along with other statistics-based metrics) could be used as an indicator of the effects of increasing damage instances but not as a direct indicator of the progression of damage. Researchers have also attempted to make data-driven predictions about the locality of the damages using EMI measurements recorded with the exponential law of the measured attenuation \citep{cherrier_damage_2013,pommier-budinger_damage_2013}. These approaches demand the use of multiple transducers for triangulation where the proximity of the sensor to the damage plays an important role. The RMSD and XCORR-based metrics are effective for stage 1 but may not be the best option to get past stage 1. There is a need to move towards a more informative approach to damage detection that is formulated with a focus on the underlying physical model of these systems. By tapping into the richness of the available data, one would be equipped with the ability to draw remarks about the severity and/or types of defects. This ultimately would help the users move efficiently along the aforesaid stages of SHM.

From the perspective of effective health monitoring, a more favorable technique would provide information about the underlying system, such that it assists in advancing along the SHM stages. 
EMI measurements are generally recorded at high-frequency ranges, typically 30-400 kHz \citep{park_overview_2003}. The high-frequency nature of these measurements makes damage characterization a challenge. The complexity of information, the amount of information required to analyze, or the challenges faced in processing all the information increases multi-fold as one moves towards stage 5 of SHM i.e., the prognosis of the damage. The previously discussed data-driven approaches fall short of providing intuition for changes due to damage. The approach presented in this work is based on physical aspects of the structure and is proposed as a platform from which to move forward onto the next stages of SHM. Influence of temperature on the EMI measurement has been showcased through various studies \citep{davoudi_effect_2012,huynh_quantification_2017}. The effects of temperature can be incorporated along the proposed approach with no negative effect as it is based on EMI measurement processing.


\section{Pole based approach}
\label{sec:Polebasedapproach} 
\begin{figure}[ht]
        \centering
        \includegraphics[width = 0.6\textwidth]{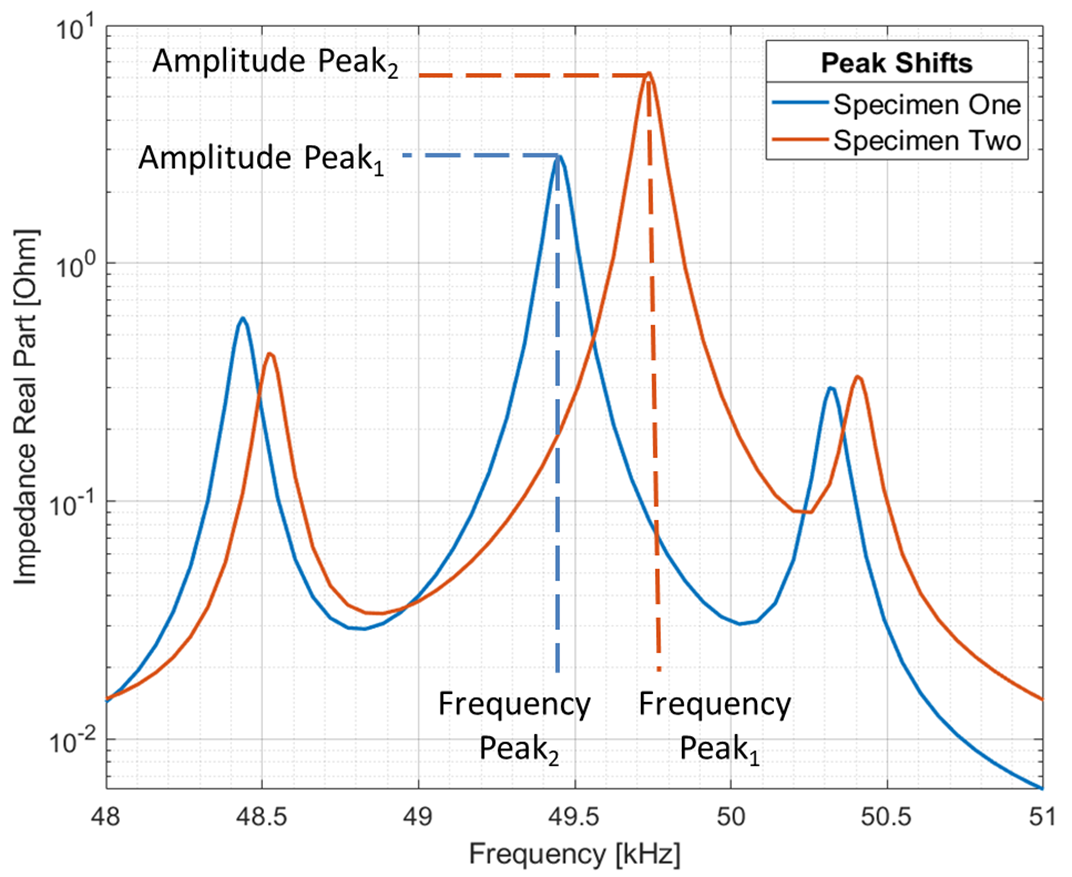}
        \caption{Estimated modal parameters of two structures to identify changes.}
        \label{fig:modalparams}
\end{figure}

Advancements in the health monitoring practice can be achieved once the potential challenges related to frequency range selection, and information averaging, are addressed. One needs a metric that does not average out the information for the complete frequency range, second, can ease window size or location requirements, and third moves the user to improved physical insights about the underlying system. It is proposed that transitioning towards metrics that consider the changes in modal parameters (or the poles of the underlying system) can assist in addressing these challenges. Identifying and understanding the modal parameters would help one disregard bands on the frequency range that do not have critical information and put emphasis on regions rich with information.
Modal parameters are sensitive to structural changes and are linked to the mass, stiffness, and damping matrices of the underlying system's state space. Hence, they are well-equipped to provide information about the system in terms of its physical attributes. Figure \ref{fig:modalparams} shows simulated EMI measurements for two specimens over a randomly selected frequency range. For a selected range, three modal peaks can be identified for both specimens. An overall rightward shift can be observed in the EMI measurement for Specimen Two. From a fundamental perspective, the natural frequency of any mechanical system is directly proportional to the stiffness and inversely proportional to the mass of the system. Thus, rudimentary inferences about physical changes between the two specimens or two distinct measurements of a specimen recorded for a pristine and damaged state can be drawn by understanding the changes in modal parameters (obtained from estimated poles). It is worth noting that the authors are focusing on poles and not necessarily EMI measurement "peaks" which can be present i.e.,  due to experimental noise and misaligned sensors and not necessarily be linked to structural changes. This differs inherently from "peak picking" techniques \cite{ogorman_converging_1983} as applied to EMI. 

The field of vibrations is based on working with the modal parameters of a structure's underlying system - natural frequencies, damping, and mode shapes. Any system's transfer function can be recreated using these modal parameters. Two systems can be distinguished from one another based on these inherent parameters. \citet{doebling_damage_1996} presented a review of studies wherein the modal parameters were identified as damage-sensitive features that can be utilized for detection. However, the literature discussed in the review deals with frequency ranges of 1000 Hz or less. 
Since modal parameters are sensitive to structural changes, for instance, an introduction of damage; distinguishing between a healthy and a damaged structure becomes intuitive. The proposed approach follows a similar intuition for damage detection wherein the modal parameters are compared to qualify the state of a structure. The modal parameters are estimated based on high-frequency EMI measurement and then used for better-informed damage detection. Thus, differences or variations in the modal parameters of the undamaged baseline and the damaged structure can be used to distinguish between the structures.

Over the last couple of decades, in the field of modal analysis, many frequency response-based modal parameter estimation algorithms have been developed. \citet{allemang_unified_1998} presented an approach to highlight the similarities between commonly used modal parameter estimation algorithms. In cases where the finite element/state space models are either too sophisticated or unavailable, one can turn to data-driven models. Existing input/output information in the form of transfer functions is used to build these data-driven models. A transfer function, $\mathbf{H(s)}$, as a function of poles and residues can be written as,
\begin{equation}
    \mathbf{H(s)} = \frac{Y(s)}{U(s)} = \sum^{N}_{i=1} \frac{r_{i}}{s - \lambda_{i}}
    \label{eq:tfaspoleresidue}
\end{equation}where $U(s)$, $Y(s)$ are the input-output functions, respectively, while $\lambda_{i}$, $r_{i}$ are the poles-residues of the underlying system.

A change in the values of the poles causes a change in the associated transfer function, and vice-versa. The poles are related to the modal parameters of the structure. By identifying the changes in the poles of a transfer function, the changes in the modal parameters of the associated structure can be investigated. An EMI measurement is a complex-valued transfer function between a complex voltage and current input-output pair. As established in the work of EMI-based SHM, any change in the physical properties of the underlying system affects the EMI measurement i.e., the transfer function. Thus, a change in the physical properties in a most basic sense would translate to a change in the poles of the input-output transfer function. By identifying changes in the poles of the underlying systems (two comparative specimens or a baseline and consecutive measurements), remarks about the "damage" can be made. Now that the problem of health monitoring has been connected to pole estimation, the next step would be choosing an approach to estimate or identify the poles. 

Many of the classical modal identification algorithms mentioned by \citet{allemang_unified_1998} are in fact mathematics-based data-driven system identification methods. Fundamentally, every algorithm builds a low-order mathematical model to represent the complex multiple-degrees-of-freedom system. This is achieved by either approximation, interpolation, or a combination of both. One of the common characteristics of these algorithms is the approximation of the transfer function either in barycentric form or approximation of numerator and denominators, independently. These approximated transfer functions undergo an iterative process such that the least square error between the approximated and the actual transfer function is minimized. 


\subsection{Pole estimation via Vector Fitting} \label{subsec:polemethod}
\begin{figure}[h!]
    \centering

        \includegraphics[width = 1\textwidth]{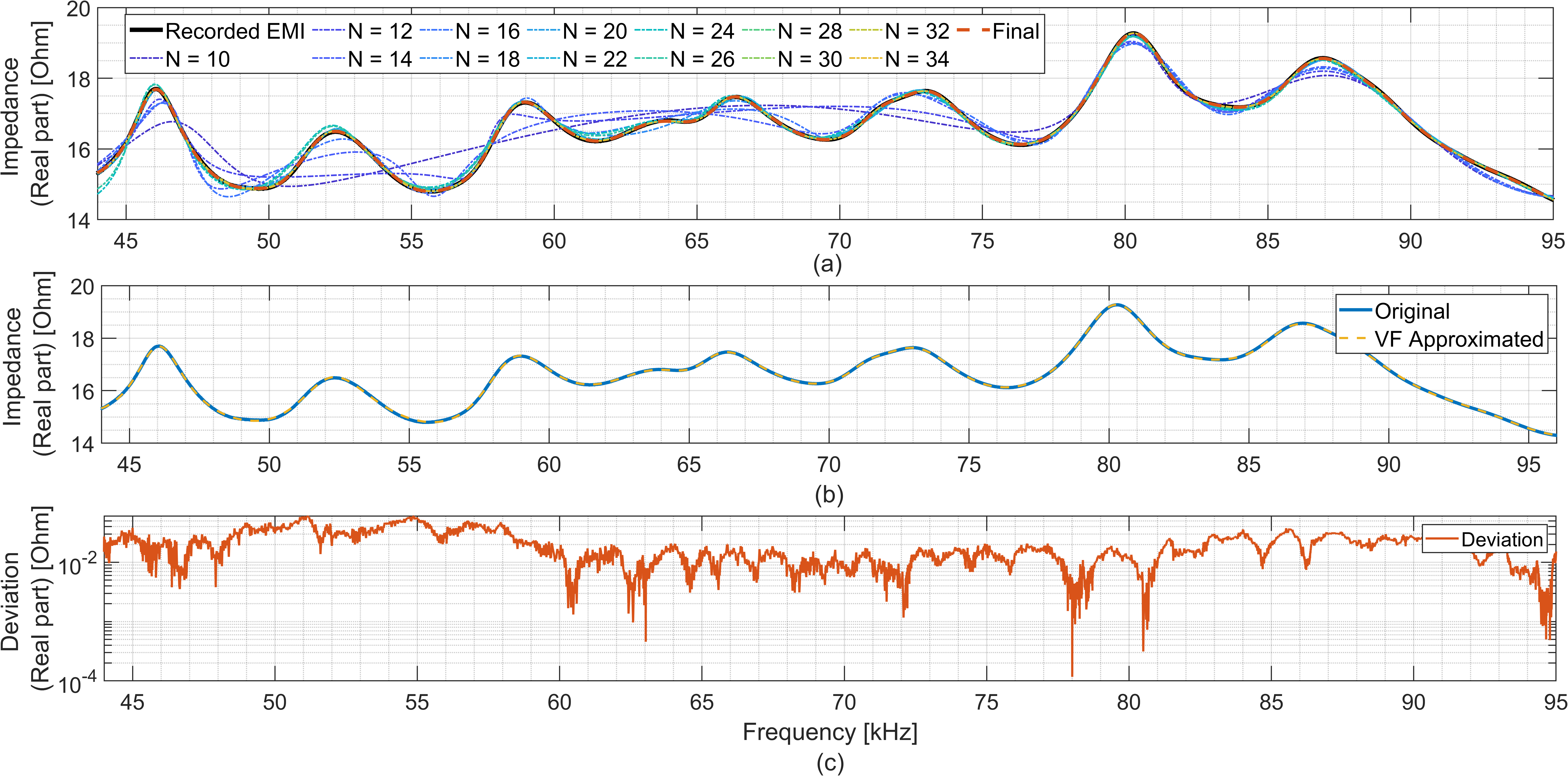}
        \caption{(a) Recorded original EMI with VF approximation for pole order 10-34, (b) Original and VF approximated EMI measurements, and (c) Deviation between original and approximated}
        \label{fig:Vfexperimenatl_dev}
\end{figure}

Parameter estimation through modeling can be cumbersome when dealing with high-frequency domains of interest. For instance, a high-fidelity finite element model can be used to simulate the high-frequency EMI measurements discussed herein and further processed to estimate the model parameters. However, the use of conventional finite element models is limited due to the high computational cost and large number of parameters required for an accurate model. \citet{albakri_dynamic_2017} discussed the use of the spectral element model to simulate EMI measurements. Even in such cases, parts remained fairly simple in geometry. For recorded as well as simulated high-frequency measurements, using model reduction techniques to estimate the responses of the underlying dynamical system can help tackle the issue of prohibitive computational costs. Such reduced-ordered models have lower dimensions than the original system, are fast to solve, and can approximate the underlying dynamics effectively. For this work, the authors leverage Vector Fitting (VF) \citep{gustavsen_rational_1999,gustavsen_improving_2006,deschrijver_macromodeling_2008}, a model-order reduction technique widely adopted in the electrical engineering domain. VF is used to approximate the EMI measurement transfer function and further identify the poles of the underlying system. This algorithm approximates the transfer function, $\mathbf{H}(s)$, over $M$ frequency points in the form,
\begin{equation}
\mathbf{H}(s) \approx \hat{\mathbf{H}}(s) = \sum^{N}_{n=1}\frac{\mathbf{c}_{n}}{s - \mathbf{a}_{n}} + d + sh \hspace{50pt}  s \in [s_1, s_2, \dots s_M],
    \label{eq:VectFit}
\end{equation}
where $\mathbf{\hat{H}}(s)$ is the estimated transfer function, $\mathbf{a}_{n}$'s, and $\mathbf{c}_{n}$'s are the poles and the residues of the approximated function, respectively, while $d$ \& $h$ are real-valued optional terms used to enforce asymptotic convergence. The order of approximation, $N$, is user-defined. The poles and residues are either real quantities or come in complex conjugate pairs. However, it is recommended to have complex starting poles linearly distributed over the frequency range of interest. VF solves the non-linear problem in Equation \eqref{eq:VectFit} sequentially as linear problems in two stages. 
The barycentric form of the transfer function is formulated as,
\begin{subequations}
\begin{equation}
    \mathbf{p}(s) = \hat{\mathbf{H}}(s)\sigma(s) =   \sum^{N}_{n=1}\frac{{c}_{n}}{s - {q}_{n}} + d + {s}h = h \frac{\prod (s - {z}_{n})}{\prod (s -{q}_{n})},
    \label{eq:VectFit1}
\end{equation}

\begin{equation}
    \sigma(s) = \sum^{N}_{n=1}\frac{\tilde{{c}}_{n}}{s - {q}_{n}} + 1 = \frac{\prod (s - \tilde{{z}}_{n})}{\prod (s - {q}_{n})},
    \label{eq:VectFit2}
\end{equation}

\begin{equation}
    \hat{\mathbf{H}}(s) = \frac{p(s)}{\sigma(s)} = h \frac{\prod (s - {z}_{n})}{\prod (s - \tilde{{z}}_{n})},
    \label{eq:VectFit3}
\end{equation}
\label{eq:VectFit1-3}
\end{subequations}
where $\mathbf{q}_{n}$'s are the $\mathbf{N}$ starting poles while $\tilde{c}_n$'s are the residues of $\sigma(s)$. It is to be noted that $\sigma(s)$ and $\hat{\mathbf{H}}(s)$ have the same poles. The ambiguity of $\sigma(s)$ is removed by forcing it to reach unity. From the barycentric form in Equation \eqref{eq:VectFit1-3}, it can be seen that the poles of the approximated transfer function are equal to the zeros of $\sigma(s)$. To obtain the zeroes of $\sigma(s)$, the corresponding residues are to be calculated. Using Equation \eqref{eq:VectFit1-3}, a linear problem can be constructed as,

\begin{equation}
    \sum^{N}_{n=1}\frac{{c}_{n}}{s - {q}_{n}} + d + {s}h = (\sum^{N}_{n=1}\frac{\tilde{{c}}_{n}}{s - {q}_{n}} + 1) \hat{\mathbf{H}}(s)
    \label{eq:VectFit_LS}
\end{equation}
Thus, an eigenvalue problem is formulated as,

\begin{subequations}
\begin{equation}
   \mathbf{A} \cdot \mathbf{x} = \mathbf{b} \hspace{50pt}  \mathbf{A} \in \mathbb{C}^{M \times 2(N+1)}, \hspace{10pt} \mathbf{x} \in \mathbb{C}^{2(N+1) \times 1} \hspace{10pt}, \mathbf{b} \in \mathbb{C}^{M \times 1},
   \label{eq:VectFit_eig1}
\end{equation}

\begin{equation}
\begin{split}
    \mathbf{A} = \begin{bmatrix}
                    \frac{1}{(s_{1} - q_{1})} & \dots & \frac{1}{(s_{1} - q_{N})} & 1 & s_{1} & -\frac{{H}(s_{1})}{(s_{1} - q_{1})} & \dots & -\frac{H(s_{1})} {(s_{1} - q_{N})} \\
                    \frac{1}{s_{2} - q_{1}} & \dots &                                  &   & s_{2} &                                             & \dots & -\frac{{H}(s_{2})} {s_{2} - q_{N}} \\
                    \vdots                       &       &                                  &   & \ddots     &                                             &       & \vdots                                     \\
                    \frac{1}{s_{M} - q_{1}} & \dots & \frac{1}{(s_{M} - q_{N})}   & 1 & s_{s} & -\frac{{H}(s_{M})}{(s_{M} - q_{1})} & \dots & -\frac{H(s_{M})} {(s_{1} - q_{N})}
                    \end{bmatrix},
\\\\
\mathbf{x} = \begin{bmatrix}
                    c_{1} & \ldots & c_{N} & d & h & \tilde{c}_{1} & \dots & \tilde{c}_{N} 
                \end{bmatrix}^\top ,
\hspace{50pt} \mathbf{b} = H(s).
\end{split}
\end{equation}
\label{eq:VectFit_eig}
\end{subequations} 
Each row of $\mathbf{A}$ is built from the starting poles $q$. Now that residues $\tilde{c}_n$, ${c}_n$ and optional terms $d$, $h$ are obtained from Equation \eqref{eq:VectFit_eig}, the zeros of $\sigma$ can be calculated as,

\begin{equation}
    \mathbf{\hat{Z}} = eig(\mathbf{\hat{Q}} - \mathbf{\hat{b}}\cdot\mathbf{\hat{c}}),
    \label{eq:VectFit_eig2}
\end{equation} 
where $\mathbf{\hat{Z}}$'s are the zeros of $\sigma(s)$, i.e. poles of the $\hat{\mathbf{H}}(s)$, $\mathbf{\hat{Q}}$ is a diagonal matrix of $q_{n}$, $\mathbf{\hat{b}}$ is a column vector of ones and $\mathbf{\hat{c}}$ is a row vector of $\tilde{c}_n$. The calculated zeros of $\sigma(s)$ or the new poles of $\mathbf{\hat{H}}(s)$ are substituted back into Equations \eqref{eq:VectFit1-3} - \eqref{eq:VectFit_eig2} as the starting poles for the next iteration and the process is repeated for $k$ iterations or until convergence, thus, ending the stage one. The convergence can be achieved by setting a tolerance value for a root mean squared error between the approximated and original transfer functions. For the second stage, residues $\tilde{c}_n$, ${c}_n$ and optional terms $d$, $h$ are estimated using Equation \eqref{eq:VectFit1-3} and the converged poles at the end of stage one. At the end of stage two, VF provides a set of converged poles, residues, and the optional terms required to estimate the transfer function. Figure \ref{fig:Vfexperimenatl_dev} shows a basic implementation of VF, wherein a recorded EMI measurement, the VF approximated measurement for various pole orders, and a deviation between the measurements (the difference between the original and approximated data) over a frequency range of 45-95 kHz are presented. It is to be noted that all calculations and approximations were performed using the complex-valued EMI. However, only the real part of the EMI has been presented throughout the work as it is more sensitive to changes than the imaginary part or the magnitude. The imaginary part and the magnitude of EMI are dominated by the capacitive response of the transducer and are less sensitive to structural changes \citet{sun_truss_1995}.

\begin{figure}[h!]
    \centering
    \includegraphics[width = 0.95\textwidth]{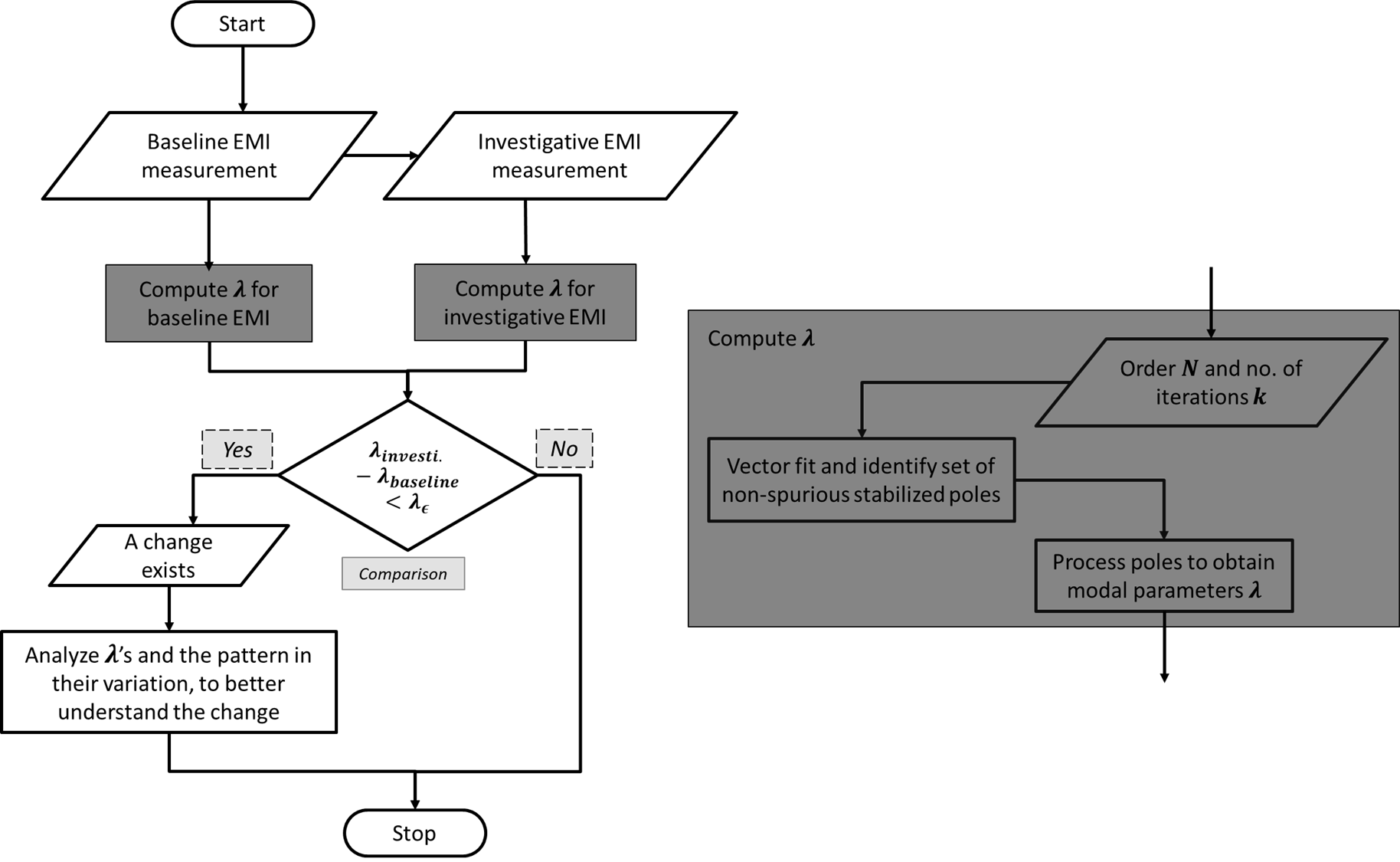}
    \caption{Process flow for damage detection using Pole estimation via VF. Modal parameters '$\lambda$' refer to the estimated modal frequencies and damping ratios of the system.}
    \label{chart:SHM_VF_flowchart}
\end{figure}

\pagebreak
The proposed VF-based approach for damage detection is summarized in Figure \ref{chart:SHM_VF_flowchart}. Initially, a baseline EMI measurement is recorded. A baseline measurements refers to measurement of a structure in a pristine (undamaged) state. This measurement is then approximated using VF, for a user-defined order ($N$) and a number of iterations ($k$). To gauge the appropriate number of poles needed to estimate the measurement, the authors recommend incrementally changing $N$ to estimate the measurement and by extension the system poles. The motivation behind this is to distinguish between physical and spurious modes by post-processing the obtained poles. As $N$ is increased, the physical poles are stabilized and located within a tolerance of variation. Once the stabilized set of poles (i.e., non-spurious poles) is identified, one can compute the modal frequencies and the corresponding damping characteristics. This procedure is analogous to building a stabilization diagram commonly used in modal parameter estimation \citep{van_der_auweraer_discriminating_2004}. Further, an investigative EMI measurement is recorded. This refers to a second measurement for the original structure after a suspected damage has occurred. Similar to the baseline measurement, a set of stabilized poles is estimated for the investigative measurement. The changes in the investigative measurement when compared to the baseline measurement can be determined by observing the changes in pole locations and their corresponding modal parameters. If the variation between the modal parameters is greater than a preset tolerance value, the structure under investigation is qualified as damaged. Further, understanding the nature of variation in the modal parameters can provide more information about the damage. For example, if the modal frequencies experience an increase, conceptually, the system can be assumed to have undergone stiffening.

\newpage

\subsection{VF v. Least Square Complex Exponential} \label{subsec:LSCF}
An alternative technique to the rational function approximation is the rational polynomial estimation. To approximate the input-output relation, i.e., the transfer function, rational polynomial estimates of the numerator and denominator are generated. The Least Squares Complex Frequency domain method (LSCF) by \citet{auweraer_application_2001}, is an industry-standard rational fraction polynomial-based technique. The rational polynomial approximated transfer function has the form, 

\begin{equation}
   \hat{H}(s) = \frac{a_0 + a_1s + a_2s^2 + \dots + a_Ns^N}{b_0 + b_1s + b_2s^2 + \dots + b_Ns^N} = \frac{\sum_{i=0}^Na_{i}f_n(s)}{\sum_{i=0}^Nb_{i}f_d(s)} = \frac{\alpha(s)}{\beta(s)},
    \label{eq:tf_summa_expansion}
\end{equation}
where $\hat{H}(s)$ is the approximated response, $\alpha(s)$ is a polynomial for the numerator with the coefficients $a_{i}$ and $\beta(s)$ is a polynomial for the denominator with the coefficients $n_{i}$, while $N$ is the maximum order used for approximation. The error between the original $\mathbf{H}$ and approximated $\mathbf{\hat{H}}$ transfer functions is obtained as a difference between the values at each frequency point and is non-linear in nature. The non-linear problem is sub-optimally linearized by right multiplying with the denominator polynomial as,
\begin{subequations}
 \begin{equation}
    \epsilon_{NL} = H(s) - \hat{H(s)} = H(s) - \frac{\alpha(s)}{\beta(s)},
\end{equation}

\begin{equation}
    \epsilon_{L} = H(s)\beta(s) - \frac{\alpha(s)}{\beta(s)}\beta(s),
\end{equation}
\end{subequations}
where $H(s)$ is the original transfer function, while $\epsilon_{NL}$ and $\epsilon_{NL}$ are the non-linear and linear errors between the two transfer functions. The coefficients are obtained by minimizing the error and solving a least square error problem. Similar to VF, the order of approximation is determined by the user and followed by the generation of a stabilization diagram. Details on VF and LSCF can be found in \citep{gustavsen_rational_1999,gustavsen_improving_2006,deschrijver_macromodeling_2008,auweraer_application_2001}

\begin{figure}[h!]
    \centering
    \includegraphics[width = 0.9\textwidth]{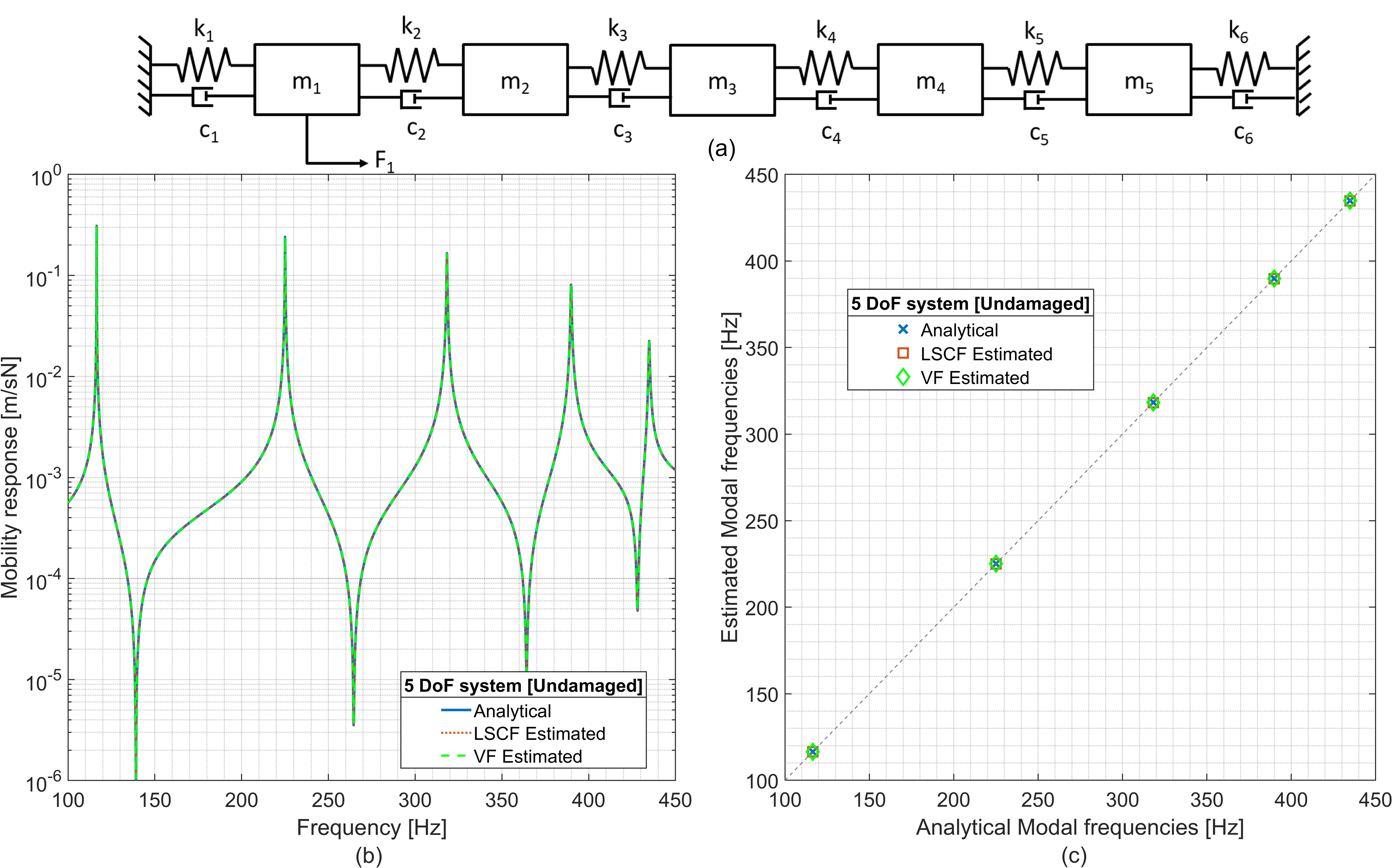}
    \caption{(a) 5-Dof simulated example for low-frequency ranges [100 - 450 Hz], (b) Analytical, VF and LSCF-estimated responses, and (c) Analytical, VF and LSCF-estimated modal frequencies. }
    \label{fig:5dof_example}
\end{figure}

To understand the comparative performances of VF and LSCF, a simulated example of a 5 degrees-of-freedom (DoF) system and an example using recorded EMI measurement are presented. These examples are representative of low-frequency and high-frequency range (operational range for EMI) applications. For low-frequency applications a discrete spring-mass-damper system, as shown in Figure \ref{fig:5dof_example}, was used. 

The operational frequency for the system was chosen to be less than 1 kHz in accordance to earlier mentioned studies wherein modal parameters were used as features for damage identification. The simulated system consists of 5 discrete equivalent masses [$m_i$ = 0.1 $Kg$] connected with springs [$k_i$ = 200$kN/m$] and with proportional damping [$c_i$ = 0.05 $N/m^2$]. Masses $m_1$ and $m_5$ are connected to the ground by springs and dampers, while a force [$F_{1} = 1$ $N$] is applied to $m_1$. To simulate damage, the stiffness, $k_2$ was reduced by $25\%$, and damping, $c_2$, was increased by $25\%$. Figure \ref{fig:5dof_example} shows the analytical drive point frequency response function corresponding to mass $m_{1}$ along with VF and LSCF approximated responses for equivalent order of approximation.

The analytical poles, modal frequencies, and damping, along with the LSCF and VF estimated ones are presented in the Appendix under Tables \ref{tab:AnlytLscfVf_undamaged} for the undamaged system and \ref{tab:AnlytLscfVf_damaged} for the damaged system, respectively. It can be observed that in the frequency range of 100 to 450 Hz, both VF and LSCF can provide good approximations for the frequency response and estimate poles and modal parameters with low error. The absolute percent difference error between the analytical and estimated modal parameters (frequency and damping), is calculated as, ($\Delta_{prms} = 100*|(Anl_{prms} - Est_{prms})/Anl_{prms})|$. For both LSCF and VF, the errors for the estimated modal frequencies were less than 0.1\% and for modal damping were close to zero. Thus, when dealing with systems or structures operating in
the lower frequency range, both VF and LSCF can be used interchangeably.

\begin{figure}[h!]
    \centering
    \includegraphics[width=0.85\textwidth]{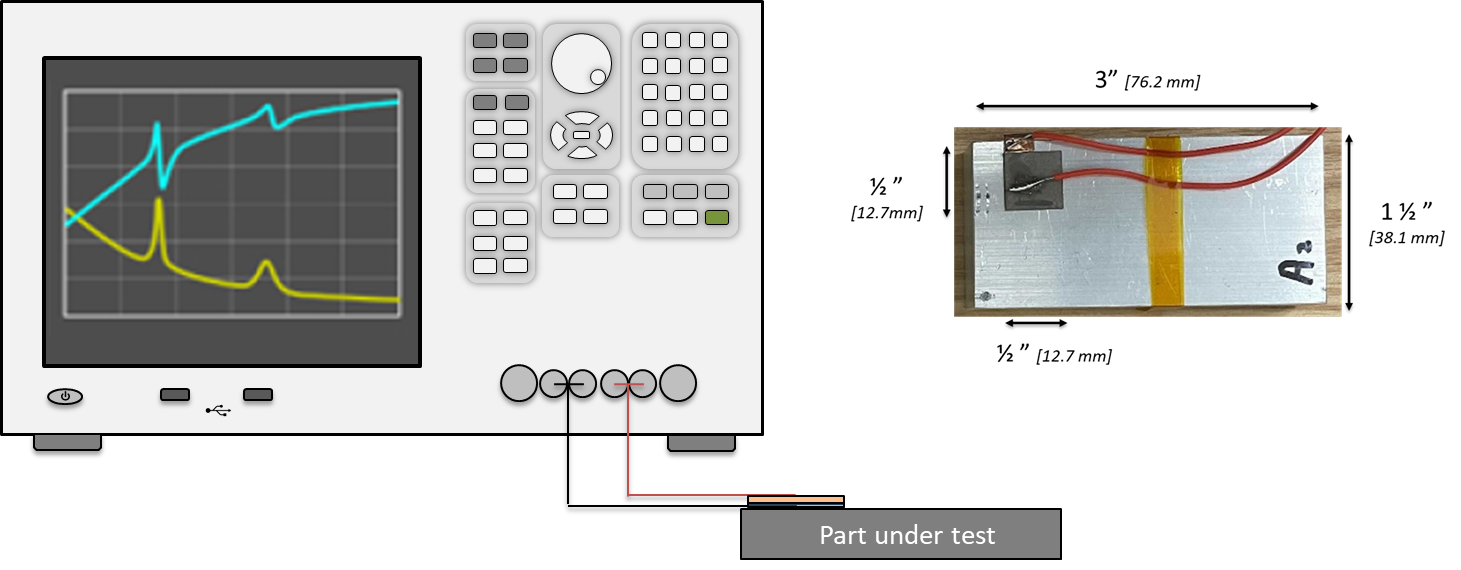}
    \caption{The experimental setup used for recording an EMI measurement and Aluminum specimen with attached piezoceramic transducer.}
    \label{fig:experimental_setup}
\end{figure}

Next, a comparison between the performances of the two techniques for high-frequency applications is carried out. An effective approximation technique for the current application would be one that can operate effectively in higher frequency ranges, i.e. EMI operative ranges. For this study, a machined aluminum block of dimensions 76.2mm x 38.1 mm x 19.05 mm was used. A monolithic piezoelectric patch (PZT 5H) of dimensions 12.7 mm x 12.7 mm x 0.2 mm was bonded to the surface of the beam using cyanoacrylate. The piezoelectric transducer was attached 2 mm from the longer edge and 6 mm from the shorter edge. Measurements were recorded over a frequency range of 940 Hz to 100 kHz. A schematic of the experimental setup used is shown in Figure \ref{fig:experimental_setup}. The specimen shown was subjected to free-free boundary conditions and excited with an incremental sine sweep applied by the Keysight E4990A impedance analyzer.

\begin{figure}[h!] 
    \centering
    \includegraphics[width=0.9\textwidth]{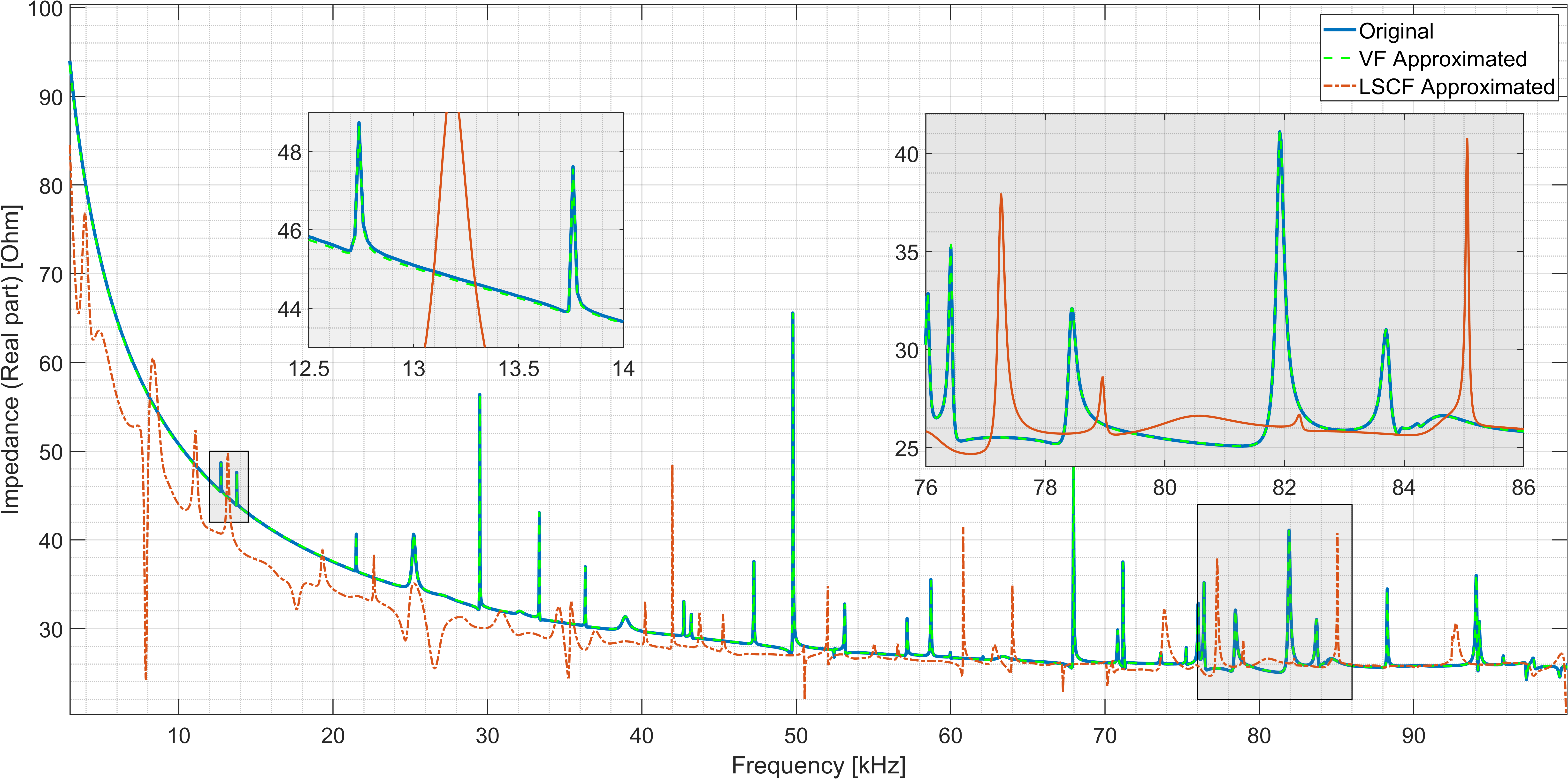}
    \caption{Original and approximated EMI measurements using VF and LSCF, over the entire range [940 Hz - 100 kHz].}
    \label{fig:Impe_lSCF_VF}
\end{figure}

\pagebreak
Initially, EMI measurements recorded for the frequency range of 940 Hz to 100 kHz, were approximated using VF and LSCF. One issue encountered with using LSCF to approximate the EMI measurement over the entire frequency range, was the difficulty in achieving convergence, as observed in Figure \ref{fig:Impe_lSCF_VF}. The least squares approximation is affected by a non-linear fitting problem. Although a right or left denominator multiplication linearizes the non-linear fitting problem, the resulting problem turns out to be badly scaled and thus, limits the use of the technique to a low order of approximation [$N$]. 

\begin{figure}[h!]
        \centering
        \includegraphics[width=1\textwidth]{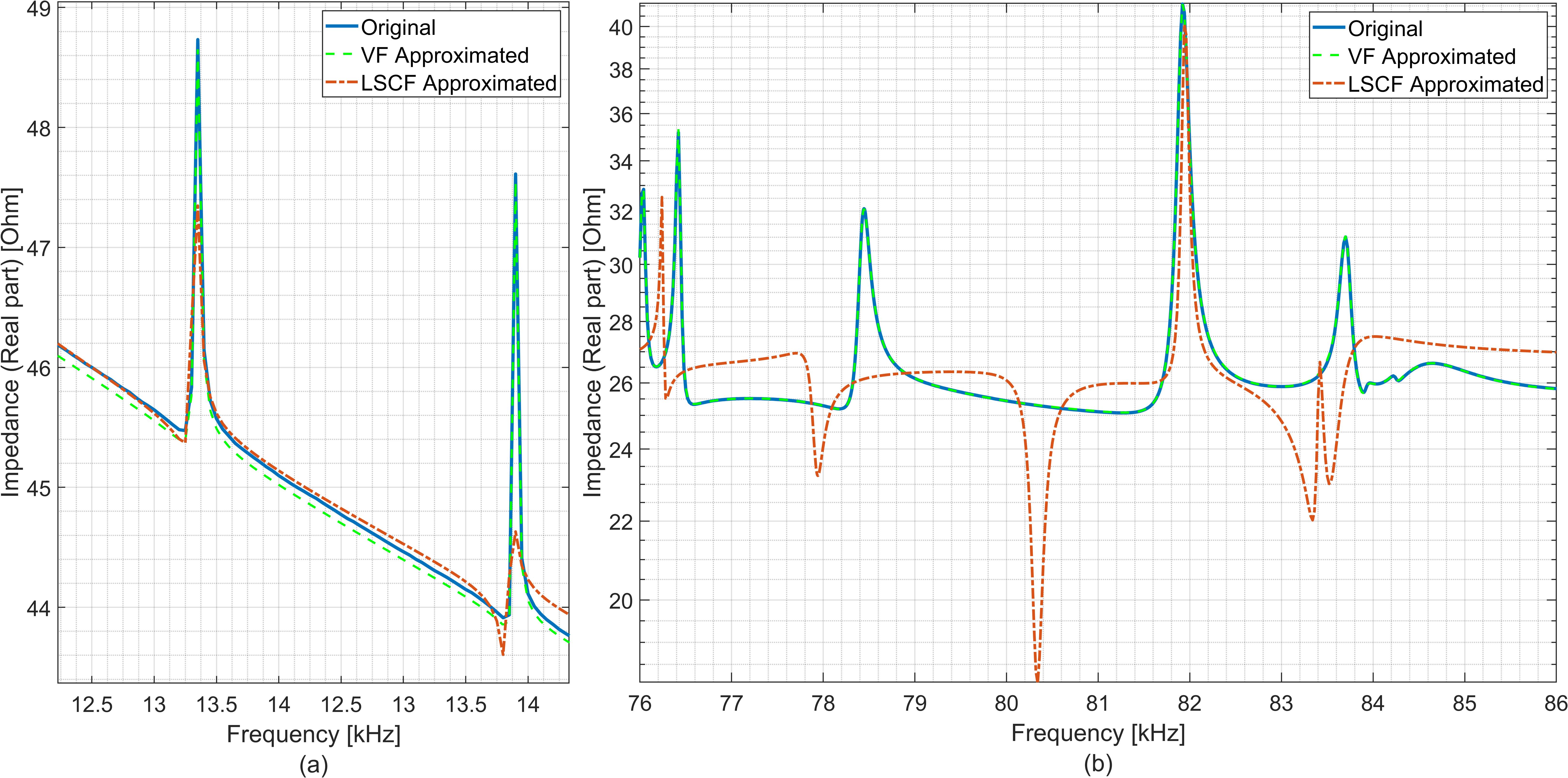}
        \caption{Original and approximated EMI using VF and LSCF, over small sections: (a) 12.5 - 14 kHz, and (b) 76 - 86 kHz}
        \label{fig:Impe_lSCF_VF_12--14_76-86}
    \end{figure}

One way to attempt to account for this limitation is to divide the total response into smaller frequency sections on which to apply the fit. As seen in Figure \ref{fig:Impe_lSCF_VF_12--14_76-86}, both VF and LSCF could approximate EMI measurements around the modal frequencies in the 12.5 - 14 kHz range. Unlike LSCF, VF was able to estimate the modal damping along with the modal frequencies in the selected range. However, the LSCF approximation struggles to match the recorded EMI measurement while VF effectively approximates the measurement in the higher frequency range of 76 - 86 kHz. 

\begin{figure}[h!]
        \centering
        \includegraphics[width=0.9\textwidth]{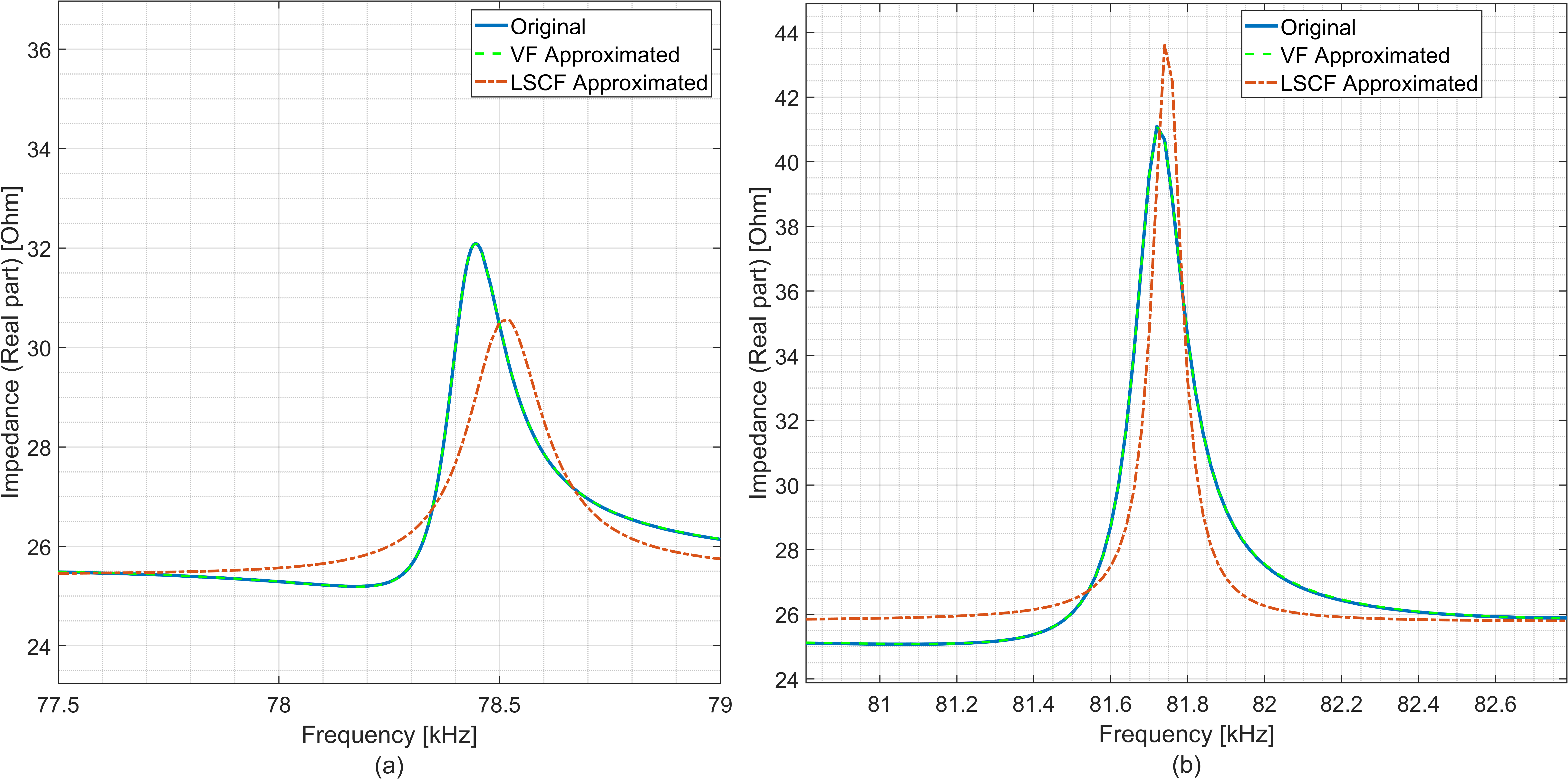}
        \caption{Original and approximated EMI using VF and LSCF, over smaller sections: (a) 77.5 - 79 kHz, and (b) 81 - 83 kHz}
        \label{fig:Impe_lSCF_VF_77-79_81-83}
    \end{figure}

\pagebreak
When trying to estimate the measurement in the higher frequency range with multiple modal peaks (large mode density), LSCF produces an even larger error. Another attempt was made to approximate individual modal peaks in the higher frequency range using LSCF, by dividing the measurement into even smaller sections to include a single modal peak. Figure \ref{fig:Impe_lSCF_VF_77-79_81-83} shows recorded measurements along with approximations around modal peaks of 78 kHz \& 82 kHz. When approximating over the smaller frequency windows, LSCF provided a better estimation than approximating over a larger or complete range. Although dividing the total frequency range into smaller sections can provide an improved LSCF approximated measurement, the process is cumbersome and produces approximation with an error higher than VF. On the other hand, it was fairly straightforward to generate a VF-approximated EMI measurement over the complete frequency range (and sections as needed), with high levels of accuracy. Thus, VF does have an advantage when compared against LSCF, to estimate high-frequency EMI measurements. 

\newpage

\subsection{VF v. Selected model reduction techniques} \label{subsec:AAA_RK}

\begin{figure}[h]
        \centering
        \includegraphics[width = 1\textwidth]{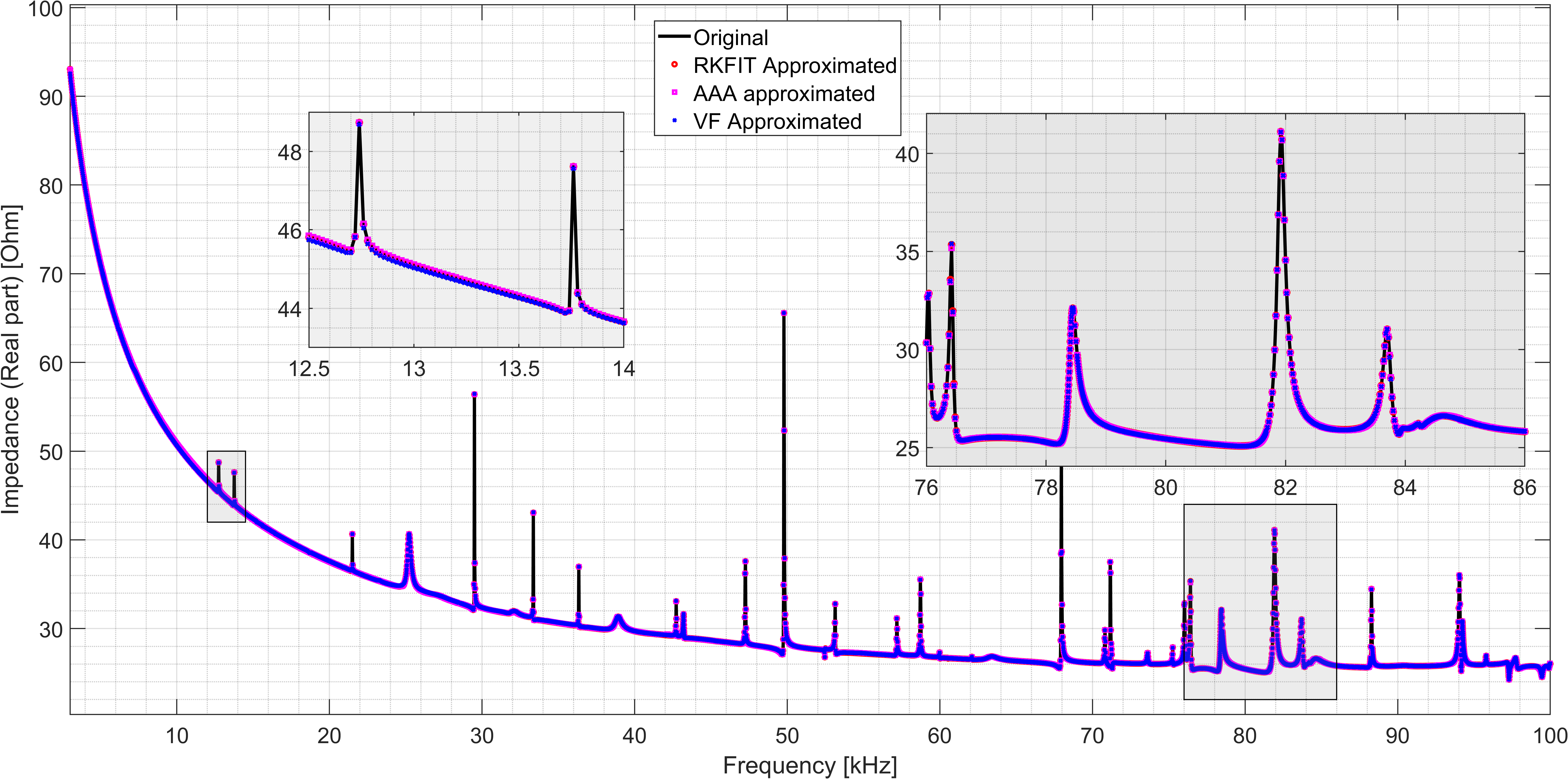}
        \caption{Original and approximated EMI measurements using RKFIT, AAA, and VF, over the entire range [940 Hz - 100 kHz]}
        \label{fig:impe_rk_aaa_vf}
\end{figure}

In Section \ref{subsec:LSCF}, VF was compared to an industry-standard modal estimation technique. In this section, the comparison is extended to a few other selected techniques that are recent and applicable to this domain. From the last section, it is understood that to better interpret the EMI measurements, one needs approximation techniques that can effectively operate in higher frequency ranges. Various rational function approximation techniques can be used to approximate the EMI measurements. For this study, VF is benchmarked against \citet{berljafa_rkfit_2017}'s Rational Krylov Fitting (RKFIT), and 
\citet{nakatsukasa_aaa_2018}'s Antoulas–Anderson (AAA) techniques. A comparison between RKFIT, AAA, and VF to approximate the same high-frequency EMI measurement discussed in the previous section is presented. For comparison, approximations are generated with the same order, $N$, for all three algorithms. RKFIT and VF, unlike AAA, need a set of initial poles and a weighting function for each of the poles. RKFIT and VF were initiated using the same set of starting poles and the same weighting function. As seen in Figure \ref{fig:impe_rk_aaa_vf}, similar to VF, both RKFIT and AAA, were able to approximate the EMI measurement over the complete frequency range (the issue faced by the LSCF). 

\begin{figure}[h]
        \centering
        \includegraphics[width=0.6\textwidth]{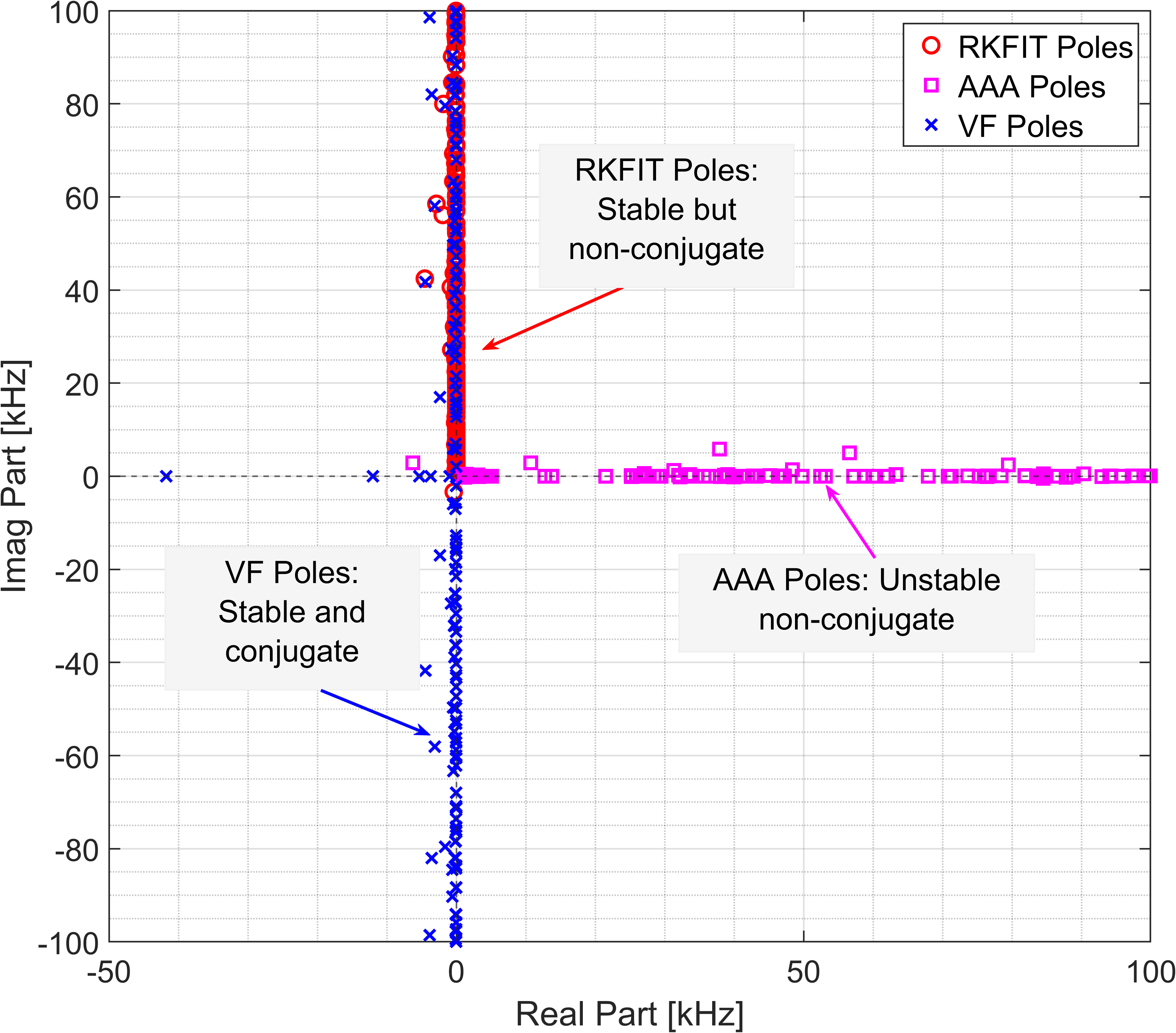}
        \caption{Poles estimated using RKFIT, AAA, and VF.}
        \label{fig:impe_rk_aaa_vf_poles}
\end{figure}

A closer look at the estimated poles yields some insight into the different methods. Figure \ref{fig:impe_rk_aaa_vf_poles} shows the poles estimated using RKFIT, AAA, and VF algorithms for the recorded EMI measurement. Although the estimated EMI measurements match the recorded EMI measurement, one can notice the differences in the poles estimated using the three techniques. The poles estimated by AAA are mostly unstable and lie in the right half plane of the plot, something that makes no physical sense for our case. While RKFIT functions similarly to VF and generates stable poles (poles in the left half plane), these poles do not occur in the form of conjugate pole pairs which are more physically appropriate. On the other hand, VF can successfully generate stable conjugate pole pairs. AAA does not require an initial set of starting poles and, thus, has advantages over the other two algorithms in terms of the minimum pre-processing but yields poles that are not physically meaningful, a goal of this work. RKFIT circumvents the issue of solving an ill-conditioned linear problem faced by VF, with the use of discrete-orthogonal rational basis functions. Approaches apart from VF are formulated with other goals in mind - better numerical precision, lower or no pre-processing data requirement, or better convergence rate. These objectives might affect the main goal of this work, which is the accuracy in modal parameter estimation with proper or interpretable physical meaning. There are ways in which the aforementioned techniques can be updated in the future to estimate poles close to what is intended, i.e. better physically relevant poles. This, however, is not in the scope of the current work. The current work provides some motivation for these potential adaptations. For this particular work, VF meets the requirements to provide an appropriate fit that matches the experimental as well as the simulated responses with precision at high-frequency ranges. Its stability and ease of use allow the user to extract the modal parameters and use them for SHM and NDE purposes. The authors believe that this will provide a more physically intuitive way of processing the EMI data and lead to the development of better metrics.

\newpage

\section{Examples for demonstration}
\label{sec:case_studies}
\quad To understand how the use of VF makes EMI-based SHM more informative and to draw its comparison with traditional RMSD-based metrics, a set of representative cases is presented here. These include simulated (Section \ref{subsec:simulated}) and experimental (Section \ref{subsec:experimental}) examples. For this work, the RMSD-based metric was chosen as the traditional metric, given its more frequent use in the field of damage detection.

\subsection{Simulated beam cases}\label{subsec:simulated}

\begin{figure}[h!]
    \centering
    \includegraphics[width = 0.75\textwidth]{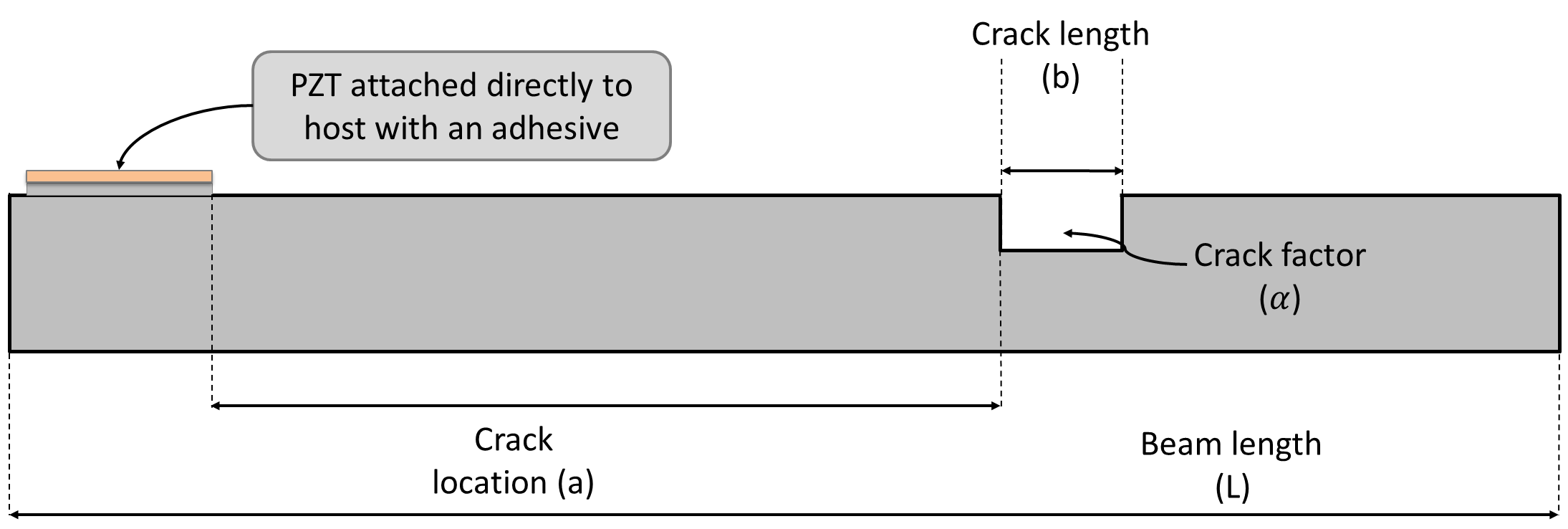}
    \caption{A schematics of the simulated crack beam with crack location (a) and crack length (b) by \citet{albakri_dynamic_2017}.}
    \label{fig:SEM_crackedbeam}
\end{figure}

A three-layered spectral element model (SEM) introduced by \citet{albakri_dynamic_2017}, was used to generate the EMI measurements of Free-Free beam bonded with a piezoceramic transducer (PZT - 5A) for seven different sets of damage parameters. The spectral element model offers an efficient solution for high-frequency simulations at a reduced computational cost compared to the finite element method-based model \citet{doyle2012wave}. Details about setting up the model and its implementation for EMI applications were thoroughly discussed in \cite{albakri_dynamic_2017} and are skipped here for brevity. The simulated beam shown in Figure \ref{fig:SEM_crackedbeam}, is an aluminum beam of dimensions 400 mm x 20 mm x 2 mm. A simulated piezoelectric sensor of size 20 mm x 20 mm x 0.2 mm, is bonded 60mm from the edge to the surface of the beam using a 20 mm x 20 mm x 0.1 mm layer of cyanoacrylate. The material properties for the simulated 3-layered beam are presented in Table \ref{tab:mat_prop}.

\begin{table}[ht]
    \centering
    \caption{Material properties for the augmented PZT-adhesive-beam system \citep{albakri_dynamic_2017}.}
        \label{tab:mat_prop}
        \scalebox{0.9}{
        \begin{tabular}{c| c c c}
        \hline 
        \multirow{2}[0]{*}{Layer} & & \multirow{2}[0]{*}{Properties} & \\ \\ \hline 

        \multirow{2}[0]{*}{Base Beam (Al 6061)} &  \multirow{2}[0]{*}{$E = 69 GPa$} & \multirow{2}[0]{*}{$\rho = 2700 kg m^{-3}$} & \multirow{2}[0]{*}{$\nu = 0.33$} \\\\
       
        \multirow{2}[0]{*}{Adhesive bonding layer} & \multirow{2}[0]{*}{$E = 1 GPa$} & \multirow{2}[0]{*}{$\rho = 1100 kg m^{-3}$} & \multirow{2}[0]{*}{$\nu = 0.36$} \\\\ 
         
        \multirow{3}[0]{*}{PZT Wafer (PZT-5A)} & \multirow{2}[0]{*}{$C_{11} = 154.13 GPa$} & \multirow{2}[0]{*}{$\rho = 7500 kg m^{-3}$} & \multirow{2}[0]{*}{$C_{66} = 25.64 GPa$} \\\\
         
        & \multirow{2}[0]{*}{$h_{31} = -8.869 x 10^8 V m^{-1}$} & \multirow{2}[0]{*}{$\beta_{33} = 1.481 x 10^8 mF^{-1}$} &  \\\\
        \hline
        \end{tabular}}
    
\end{table}

The damages were modeled as open, non-breathing cracks determined by crack location (a), length (b), and the stiffness reduction factor ($\alpha$), such that $\alpha \in$ (0,1].
The simulated EMI measurements were then used to demonstrate the aforementioned pole-based approach. A subset of the simulated damaged beam subcases summarized in Table \ref{tab:SEM_simulatedcases}, are categorized into Case I through III. The details for Case IV are provided in later sections. For Case I, a 5 mm crack with $\alpha$ = 0.6 was assumed to be located at 100mm (a) from the inner edge of the PZT. The crack length (b) was varied from 5 mm to 20 mm, to simulate progressive cases. For Case II, three independent beams with 5mm (b) cracks each with a 0.6 factor were assumed to exist at locations (a) 75mm, 100mm, and 125mm from the inner edge of the PZT. For Case III, the damage severity was changed by gradually reducing $\alpha$ from 0.75 to 0.6. All numerical simulations were performed over a 30 - 100 kHz range.

\begin{table}[h]
    \centering
    \caption{Parameters considered to simulate damages in SEM-based Free Free beam model}
    \scalebox{0.825}{
    \begin{tabular}{c c c c c}
        \hline 
         & \multirow{2}[0]{*}{Crack Location (\emph{\textbf{a}}, mm)}  & \multirow{2}[0]{*}{Crack Length (\emph{\textbf{b}}, mm)} & \multirow{2}[0]{*}{Reduction factor (\textbf{$\alpha$}, $E/E_0$)}&\\\\
        \hline 
        SW1 & 100 & 5 & 0.60 & \multirow{3}[0]{*}{Example I}\\
        SW2 & 100 & 10 & 0.60 &\\
        SW3 & 100 & 20 & 0.60 & \\
        \hline
        SL1 & 75 & 5 & 0.60 &\multirow{3}[0]{*}{Example II}\\
        SL2 & 100 & 5 & 0.60 &\\
        SL3 & 125 & 5 & 0.60 &\\
        \hline
        SF1 & 100 & 5 & 0.75 & \multirow{3}[0]{*}{Example III}\\
        SF2 & 100 & 5 & 0.65 & \\
        SF3 & 100 & 5 & 0.60 & \\
        \hline
    \end{tabular}}
    \label{tab:SEM_simulatedcases}
\end{table}

\begin{figure}[h!]
    \centering
    \includegraphics[width = 1\textwidth]{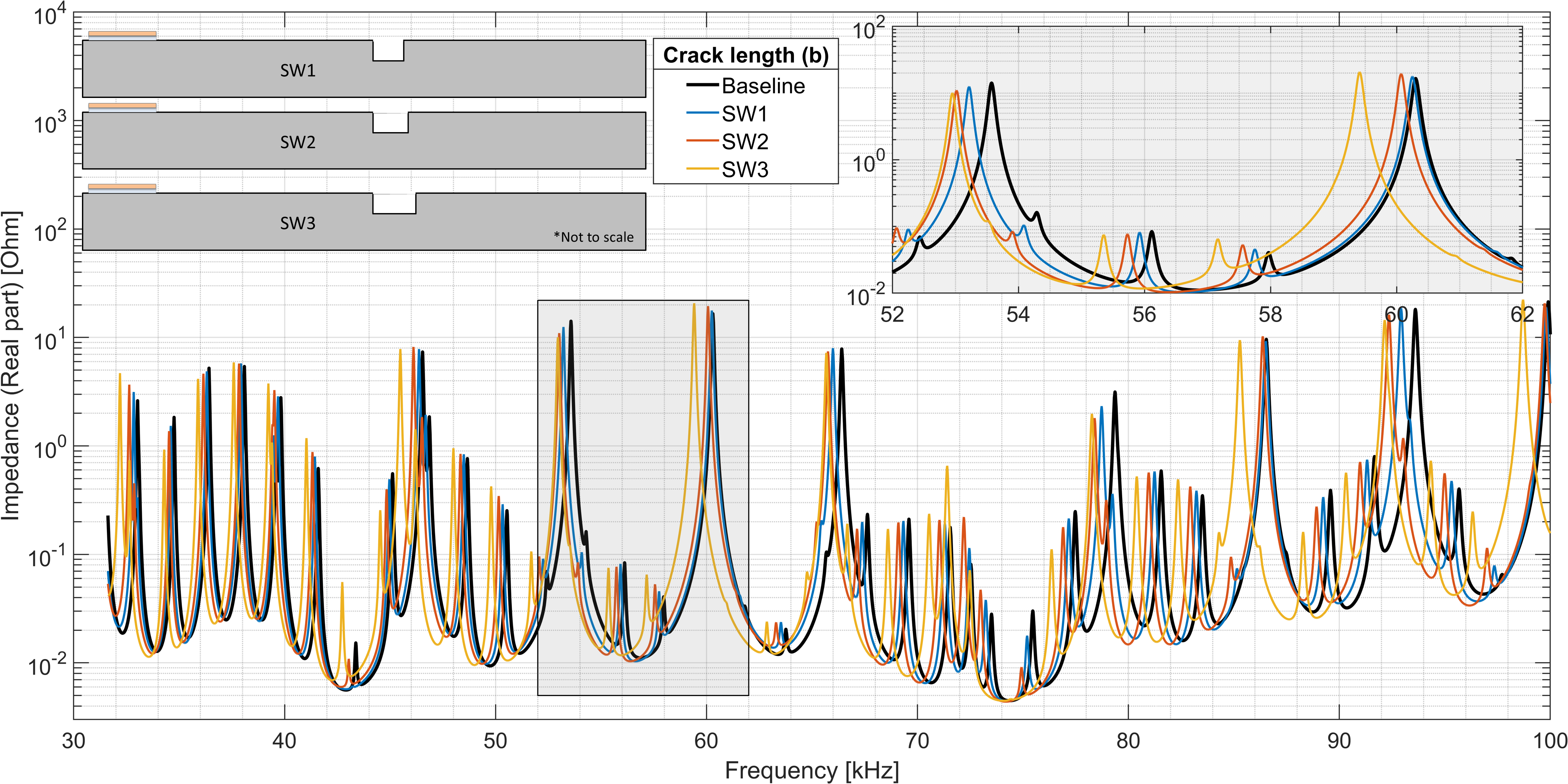}
    \caption{SEM generated EMI measurements for Case I [Zoomed over 52 - 62 kHz].}
    \label{fig:diffcracklengthzoomed}
\end{figure}

\textbf{Case I} \hspace{10mm} This is equivalent to a scenario of a propagating crack in a beam. The crack length (b) was increased from SW1 through SW3. The simulated EMI measurements, over an arbitrarily selected range of 52 - 62 kHz, are shown in Figure \ref{fig:diffcracklengthzoomed}. The amount of shift in the simulated EMI measurement varies from frequency region to region. However, in a macro-view by a user, one can observe a generalized leftward shift in the EMI measurements as the crack length increases from SW1 - SW3. 

\begin{figure}[h!]
    \centering
    \includegraphics[width =1\textwidth]{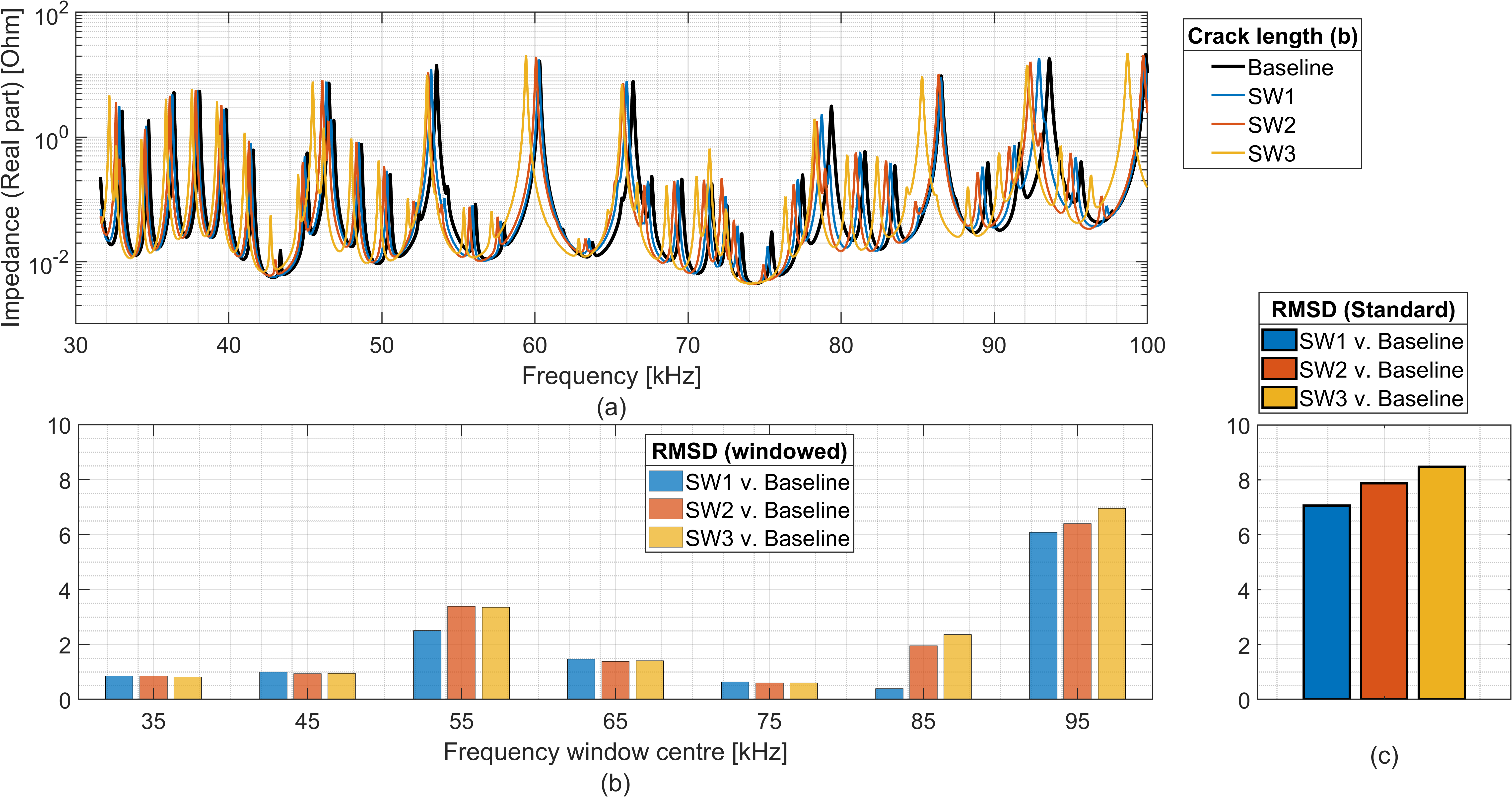}
    \caption{(a) SEM generated EMI measurements over 30 - 100 kHz, (b) Standard-RMSD metric, and (c) Windowed-RMSD metric for Case I.}
    \label{fig:diffCracklength_rmsd}
\end{figure}

\textit{Standard} RMSD-based metric (for the entire frequency range) and a windowed RMSD-based metric (for smaller consecutive 10kHz windows) were calculated for each subcase with the undamaged simulated beam as the baseline using Equation \eqref{eq:RMSD}. Henceforth, throughout this work, \textit{Standard metric} refers to the RMSD value calculated over the complete frequency range while, the \textit{windowed metric} refers to RMSD value calculated over small 10 kHz windows. The windowed metric values are plotted with respect to the center frequency value of the particular window for ease of readability. The measurements over the entire frequency range, along with the calculated metrics are shown in Figure \ref{fig:diffCracklength_rmsd}. The Standard metric values for SW1 through SW3 versus the Baseline were $7.07$, $7.87$, and $8.48$, respectively. For the frequency window of 40 - 50 kHz, the windowed metric values were $0.99$, $0.94$, and $0.96$. For the window spanning 80 - 90 kHz, the values were $0.39$, $1.95$, and $2.36$, respectively. These windowed values are not only different than the standard metric values stated earlier but also vary for each 10 kHz window. Thus, the RMSD-based metric value is affected by the frequency window of interpretation. The physical changes in the beams do not necessarily cause uniform effects in all the windows. This means that the particular damage introduced in these beams either affects only sensitive modes or has greater effects in certain windows.
The key inference from studying both the metric values is that the physical characteristics undergo gradual variation from SW1 to SW3 with the variation for SW3 being higher than that for SW1 when compared against the baseline. It is also worth mentioning that the softening effect is completely missed by using the RMSD approach.

\begin{figure}[h!]
    \centering
    \includegraphics[width = 0.6\textwidth]{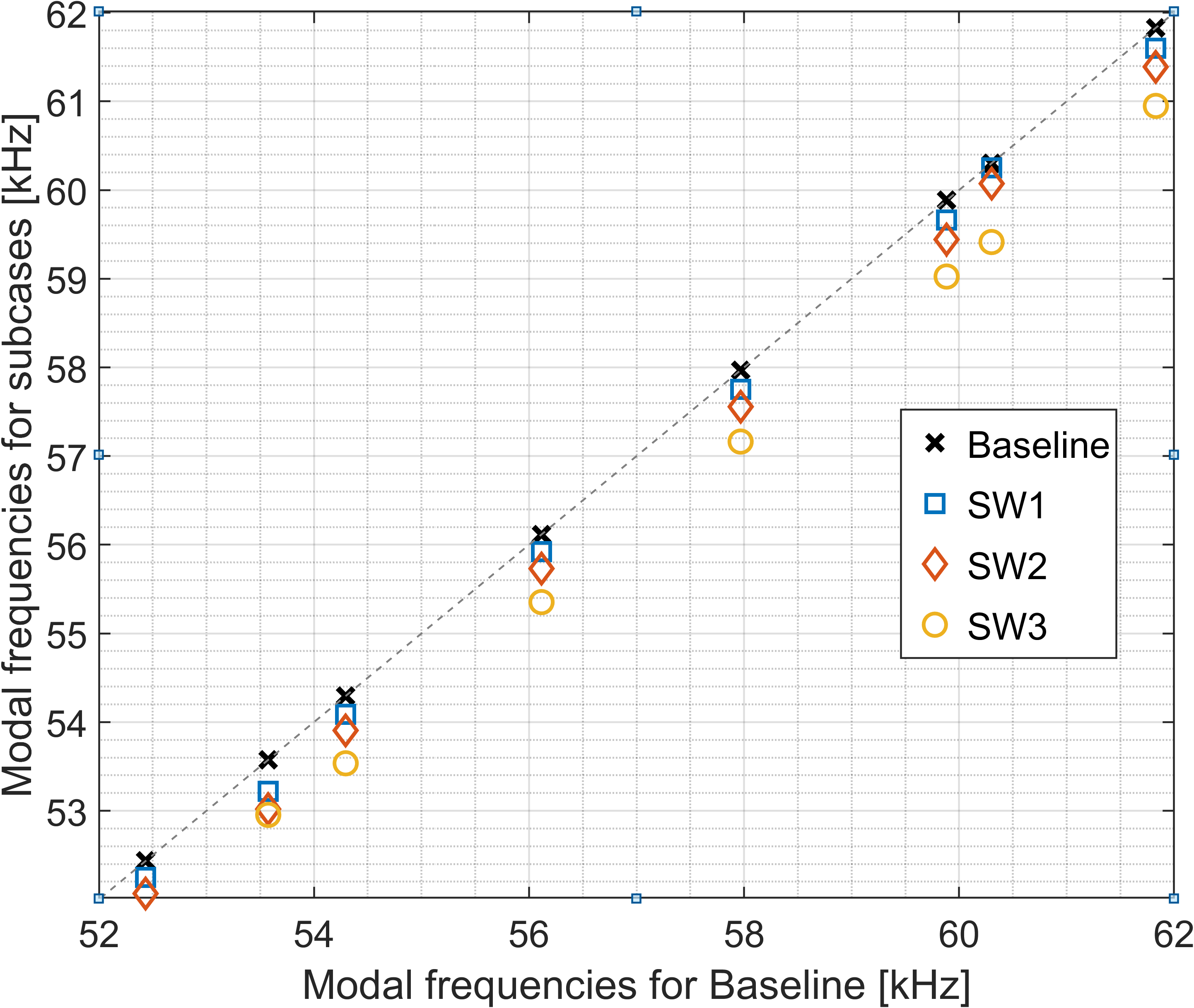}
    \caption{VF estimated modal frequencies for Case I [52 - 62 kHz].}
    \label{fig:modalfreqs_diffcracklength}
\end{figure}

\pagebreak
Next, VF was employed using the same procedure described in Section \ref{subsec:polemethod} and Figure \ref{chart:SHM_VF_flowchart} to estimate the poles from the EMI measurements. The estimated poles were then post-processed to obtain modal frequencies and damping. In Figure \ref{fig:modalfreqs_diffcracklength}, the estimated modal frequencies are plotted against the modal frequencies of the baseline over the 52 - 62 kHz range. The estimated poles, the modal parameters, and the differences between the frequencies and damping values are listed under the Appendix in Table \ref{tab:scenarioI_polestablenew}. 

\pagebreak
The modal frequencies experience a decrease in their values from SW1 to SW3. For the 52 - 62 kHz range, the modal frequencies for SW1 and SW2 experience an average reduction of about $0.37 \%$ and $0.70 \%$, respectively. For SW3, the modal frequencies experience a higher average reduction of $1.4 \%$ in their values. When looking at the bigger picture, SW3 has experienced a larger overall shift in the modal frequencies. The inference of damage effects being higher for SW3 is in agreement with the inference from studying RMSD-based metrics. However, the modal parameters provide more than just this inference. Conceptually, a general decrease in modal frequencies can imply either an increase in mass, a decrease in stiffness, or a combination of both. As no additional mass was introduced, a reduction in stiffness can be deduced. This does agree with the original premise of a propagating crack, i.e., a crack with increasing length. Thus, the changes in poles and by extension the modal parameters, provide a more qualitative understanding of the changes in the physical characteristics.

\begin{figure}[ht!]
    \centering
    \includegraphics[width=1\textwidth]{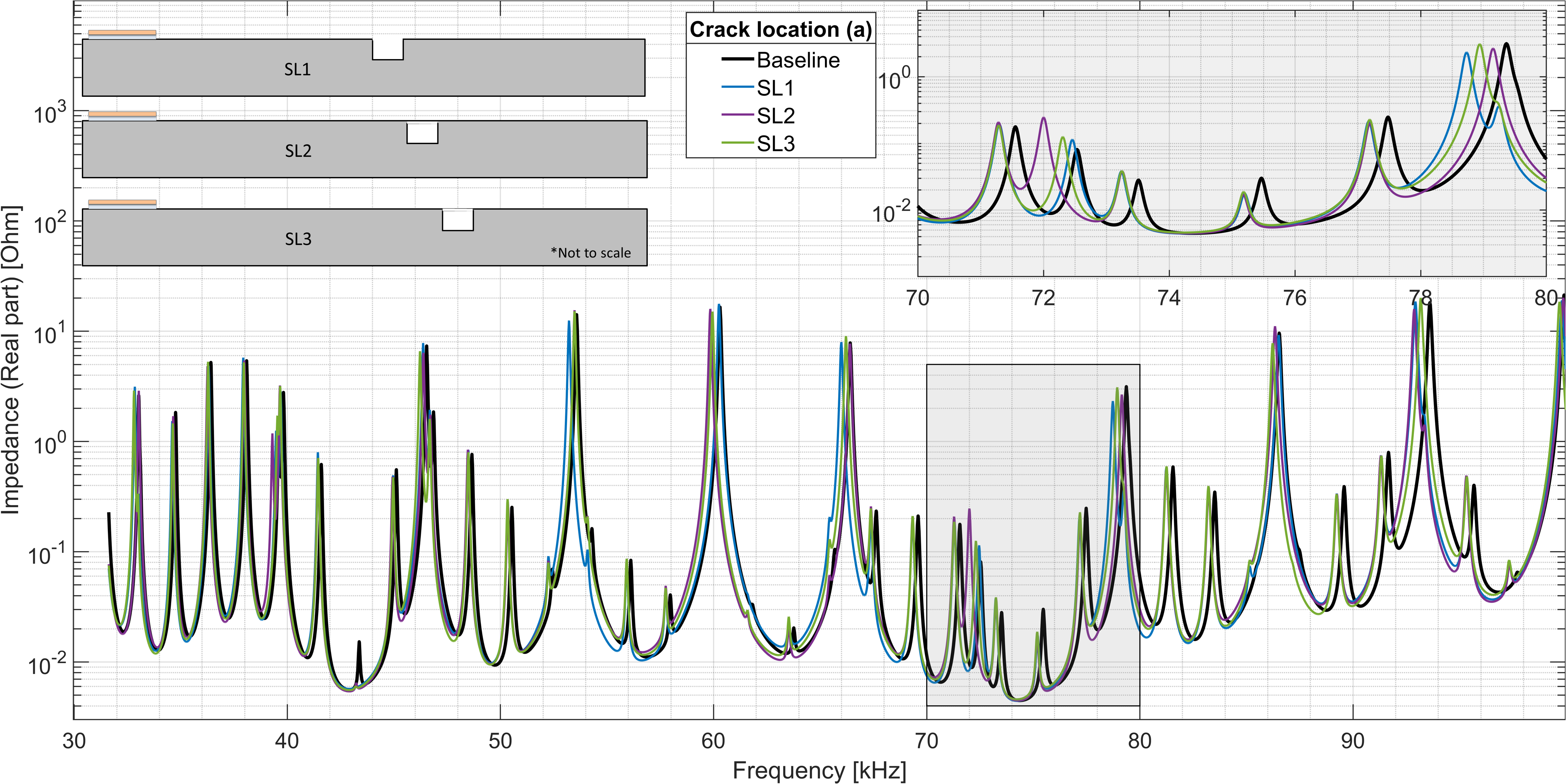}%
    \caption{SEM generated EMI measurements for Case II [Zoomed over 70 - 80 kHz].}
    \label{fig:diffcrackloczoomed}
\end{figure}

\textbf{Case II} \hspace{10mm} This corresponds to a scenario with beams with identical defects. Each beam in SL1 to SL3 has a crack with the same length (b) but a different location (a) from the inner edge of the transducer. Figure \ref{fig:diffcrackloczoomed} shows the EMI measurements for these beams, over a selected range of 70 - 80 kHz. Although the changes in the EMI measurements are evident, they do not follow an observable pattern. Unlike the previous example, the EMI measurements do not experience gradual shifts but have distinguishable regions with and without any changes. 

\begin{figure}[h!]
    \centering
    \includegraphics[width = 1\textwidth]{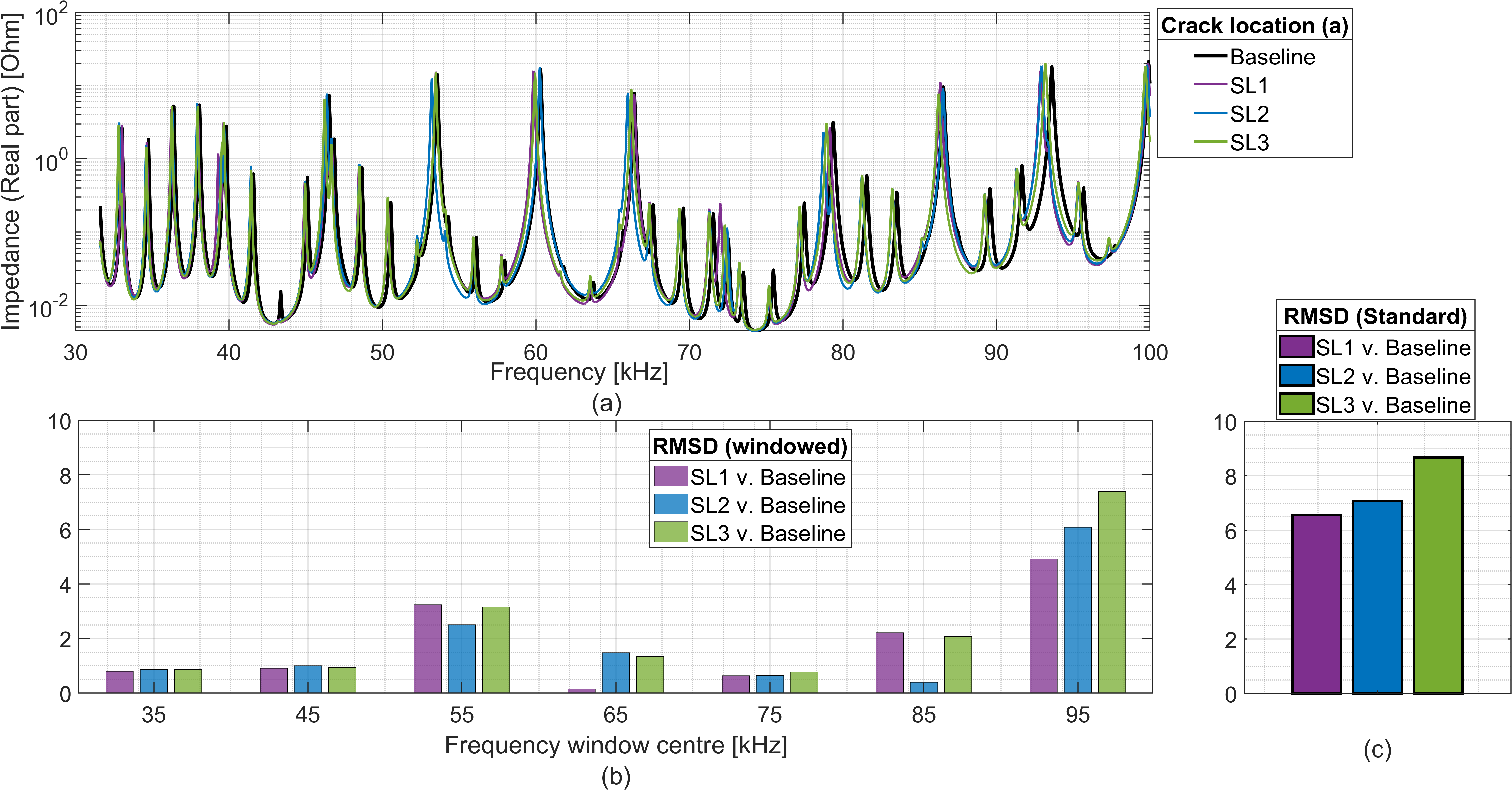}%
    \caption{(a) SEM generated EMI measurements over 30 - 100 kHz, (b) Standard-RMSD metric, and (c) Windowed-RMSD metric for Case II.}
    \label{fig:diffCrackloc_rmsd}

\end{figure}

Figure \ref{fig:diffCrackloc_rmsd} shows the EMI measurements along with the RMSD-based metrics over the entire 30-100 KHz range. The standard metrics for SL1 through SL3 were $6.55$, $7.07$, and $8.67$, respectively. For the window centered around 45 kHz, the windowed metric values were $0.91$, $0.99$, and $0.93$, while for the window spanning 50 - 60 kHz, the windowed metric values were $3.23$, $2.51$, and $3.15$. Unlike the standard metric values, the windowed metric values do not follow an overall trend. After observing the windowed metrics, one can infer that the window centered at 95 kHz has a higher value for SL3, while windows centered at 55 kHz and 85 kHz have a higher value for SL1. The windowed metric values have local patterns based on the chosen window of interpretation. However, the same conclusion does not stand true for the standard metric value as it seems to have an observable direct correlation with the relative location of the crack from the transducer. When looking at the standard metric, the values experience an increase between SL1 through SL3. Thus, one might misdiagnose this as a case of propagating crack instead of that of beams with identical cracks. The introduced defect affects the sensitive modes rather than a generic overall effect. It should be noted that this is sensitive and dependent on window sizing. This piece of information is crucial to understanding the nature of the damage. To better interpret the effects of the damage and avoid the issue of averaging out the information, one might choose to observe the windowed metric values for smaller frequency bands, but this would make the process more cumbersome.%

\begin{figure}[h!]
    \centering
    \includegraphics[width=0.6\textwidth]{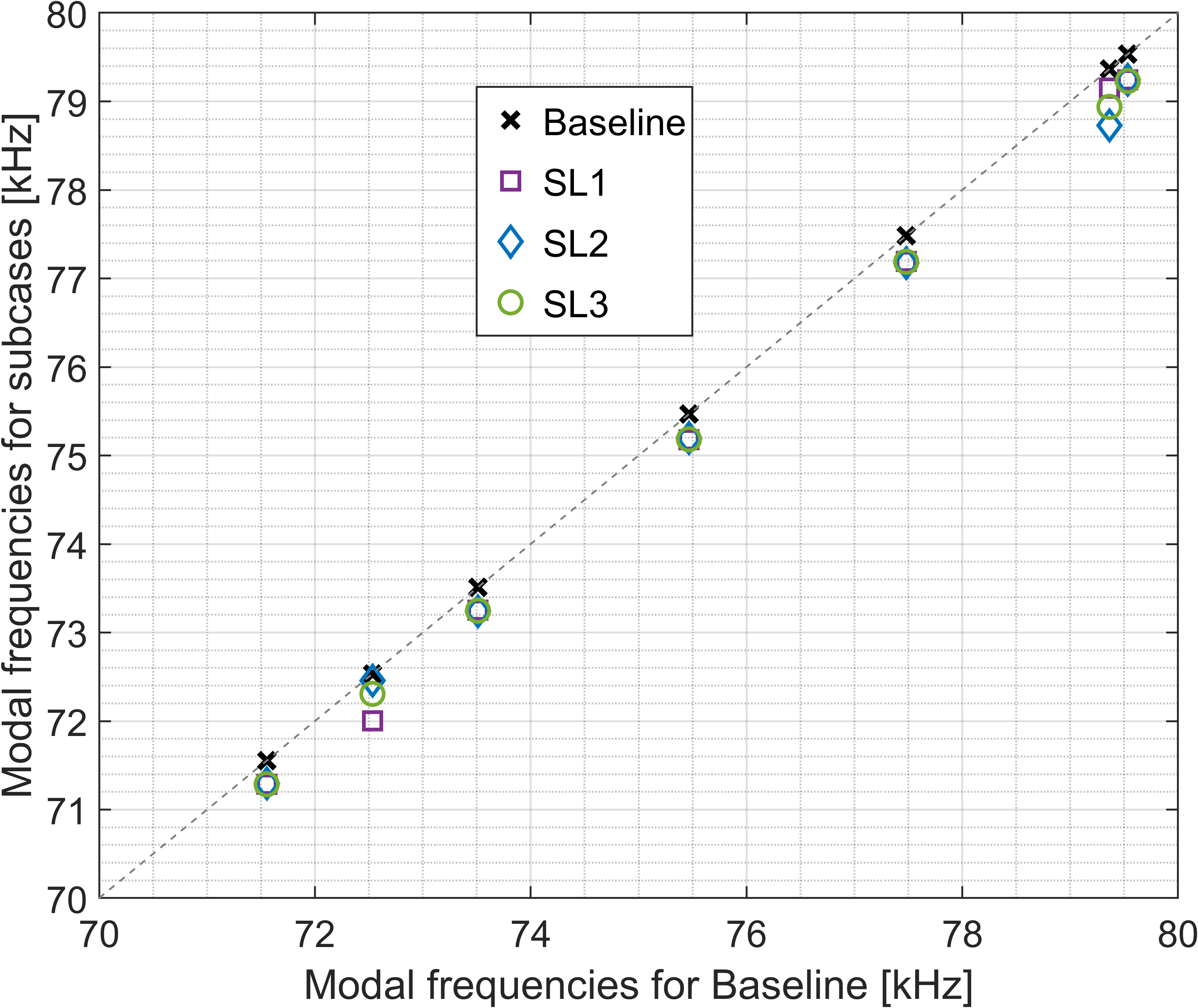}
    \caption{VF estimated modal frequencies for Case II [70 - 80 kHz].}
    \label{fig:modalfreqs_diffcracklocation}
\end{figure}

\pagebreak
However, the modal frequencies can provide some help in this aspect. The VF estimated modal frequencies are plotted against the modal frequencies of the baseline in Figure \ref{fig:modalfreqs_diffcracklocation}, while the detailed information about the estimated modal parameters is presented in Table \ref{tab:scenarioII_polestablenew}. 

\pagebreak
When observing the estimated modal frequencies, one cannot ignore the peculiar unaltered nature of most of the modal peaks. The identical nature of the damage can be one of the reasons for such behavior (although not fully addressed here). All modal frequencies for SL1 through SL3 in the chosen range, tend to be very close to one another or rather unchanged. The only exceptions to this are the modal frequencies around 72 kHz \& 79 kHz in the baseline measurement. The modal frequency around 72.5 kHz in the baseline measurement experiences a decrease of about $0.73 \%$, $0.10 \%$, and $0.31 \%$, for SL1 through SL3. However, the modal frequency around 79 kHz experiences a decrease of about $0.27 \%$, $0.80 \%$, and $0.54 \%$ for SL1 - SL3. For this example, the location of the damage causes a particular mode at ~79.2 kHz in SL1, split into significantly distinguishable two modal peaks for SL2 while preserving the single modal peak behavior for SL3 with a hint of a secondary modal peak. This indicates that the system is changing drastically and there are certain modes or windows in the frequency range more sensitive to the damage simulated for this example. RMSD-based metrics give no pattern in variation whereas VF can help identify the modes that undergo observable changes due to the defect, i.e. the sensitive modes. This points towards focusing on variations in sensitive modes, different from the overall softening effect as seen in Example I. From a wider perspective, identifying the significant modes can help one get a better idea of the wavelength of interpretation to focus on and thus, can provide a direction for improved damage detection. The authors do not claim that it is elementary to relate these inferences to particular physical changes, but these inferences do provide richer data that can help one move closer to gaining more physical information. \\

\begin{figure}[h!]
    \centering
    \includegraphics[width=1\textwidth]{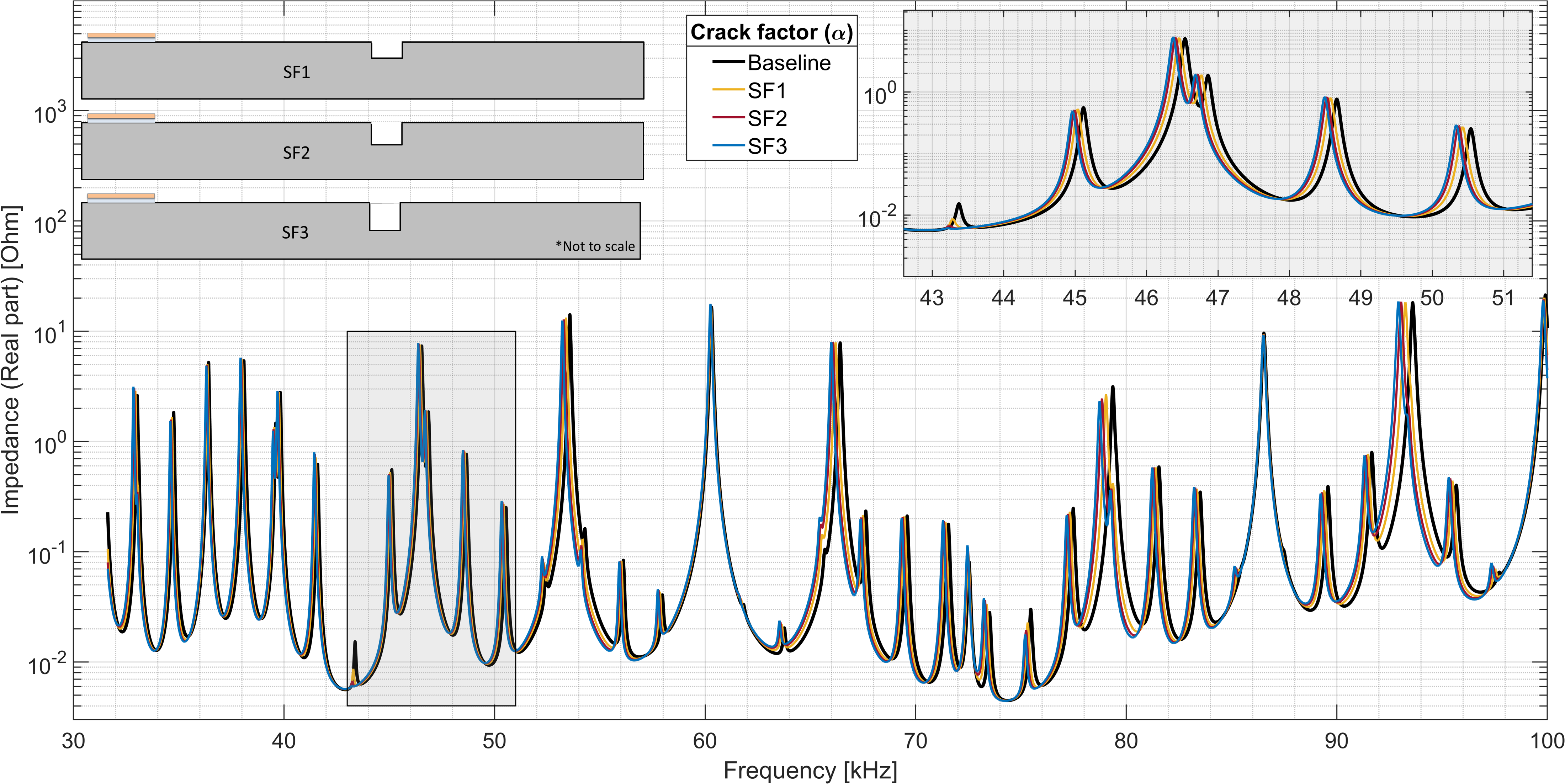}
    \caption{SEM generated EMI measurements for Case III [Zoomed over 42 - 52 kHz].}
    \label{fig:diffcrackdegzoomed}
\end{figure}

\textbf{Case III} \hspace{10mm} Another example to understand the progression of damage is that of a degrading crack (crack depth) in a beam. For this, the stiffness reduction factor \textbf{$\alpha$} was gradually decreased from SF1 to SF3. The simulated EMI measurements over 42 - 52 kHz, are shown in Figure \ref{fig:diffcrackdegzoomed}. As the $\alpha$ value is decreased to simulate the crack getting worse from SF1 - SF3, an incremental leftward shift is observed in the EMI measurements. 

\begin{figure}[h!]
    \centering
    \includegraphics[width=1\textwidth]{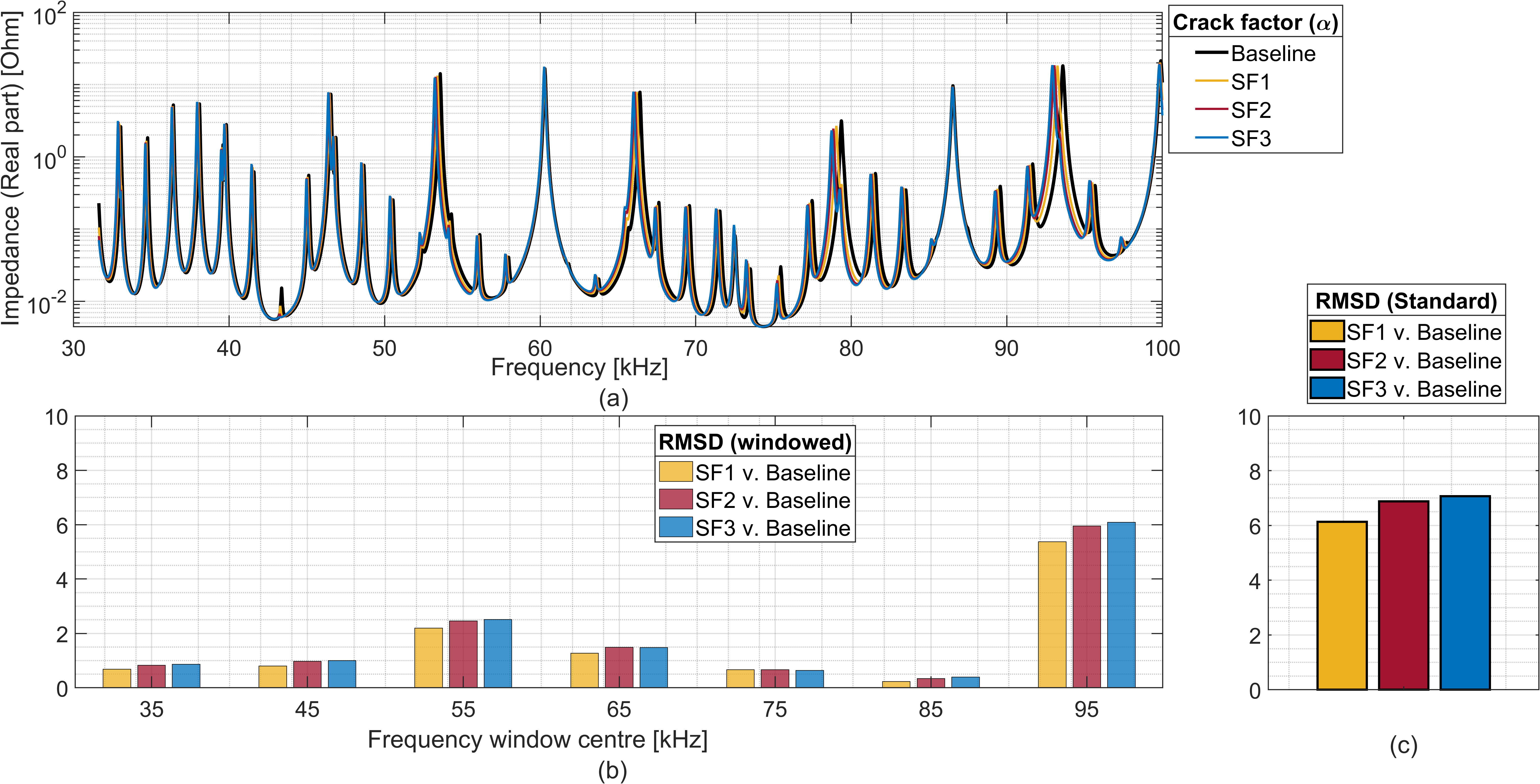}
    \caption{(a) SEM generated EMI measurements over 30 - 100 kHz, (b) Standard-RMSD metric, and (c) Windowed-RMSD metric for Case III.}
    \label{fig:diffCrackdeg_rmsd}
\end{figure}

The EMI measurements over the entire frequency range with the previously calculated metrics are shown in Figure \ref{fig:diffCrackdeg_rmsd}. For SF1 through SF3 versus baseline, standard metric values were $6.14$, $6.88$, and $7.07$, respectively. For the window spanned between 40 - 50 kHz, the windowed metric values were $0.80$, $0.97$, and $0.99$ respectively while for the window 80 - 90 kHz, the values were $0.23$, $0.34$, and $0.39$. The window of inspection affects the windowed metric value, but the overall trend of variation remains the same to some extent. As the "crack worsens", i.e. $\alpha$ decreases or the crack deepens, each of the windowed metric values experiences an increase in their values. It can be inferred that both the standard and windowed metric values experience a global incremental change with a gradual variation in the damage quality with both metric values being overall higher for SF3 than for SF1. 

\begin{figure}[h!]
    \centering
    \includegraphics[width=0.6\textwidth]{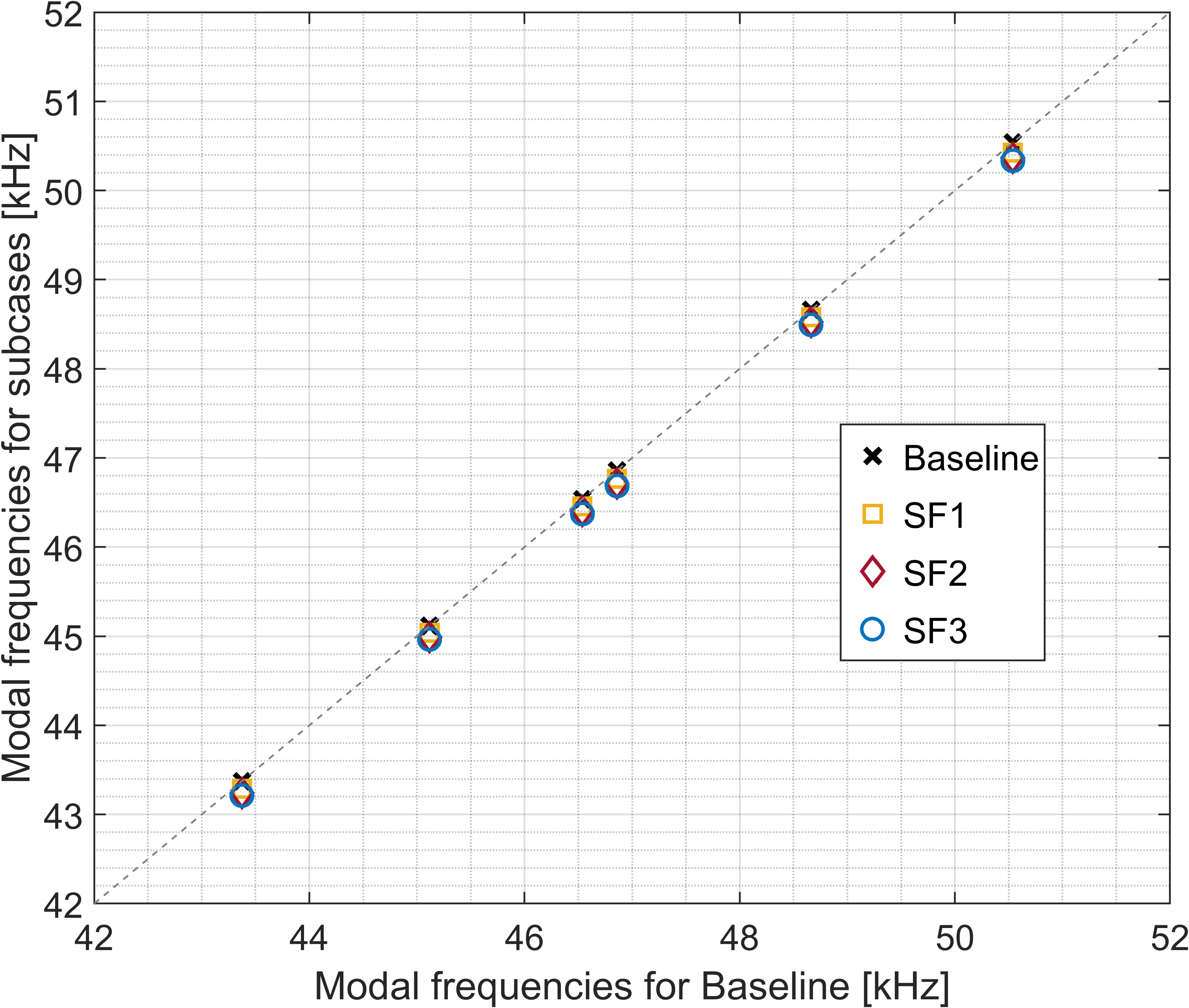}
    \caption{VF estimated modal frequencies for Case III [42 - 52 kHz].}
    \label{fig:modalfreqs_diffcrackdegree}
\end{figure}

Further, poles are estimated using VF and then processed to obtain the modal frequencies and damping. The estimated modal frequencies in the chosen frequency range of 42- 52 kHz are plotted against those of the baseline in Figure \ref{fig:modalfreqs_diffcrackdegree} and listed in Table \ref{tab:scenarioIII_polestablenew}. 

\newpage
From SF1 through SF3, the modal frequencies experience a decrease in values (easier to observe in Table \ref{tab:scenarioIII_polestablenew}). For the selected window, modal frequencies in SF3 experience a larger amount of decrease than for SF1. Similar to Case I, a global decrease in modal frequencies indicates an overall mass increase and/or stiffness reduction. Since no mass was added externally, the global defect can be inferred to be an effect of stiffness reduction. This matches the original assumption of altering the stiffness reduction factor from SF1 through SF3. Similar to cases wherein a crack length increases, an increase in crack "depth" causes a reduction in the stiffness. By observing the reduction in the modal frequencies for this case, an association with degrading crack can be established. Although both approaches provided a progressive effect in the damage, VF was able to point the user to a degrading crack under the assumption of no added mass.\\

\begin{figure}[h]
    \centering
    \includegraphics[width = 1\textwidth]{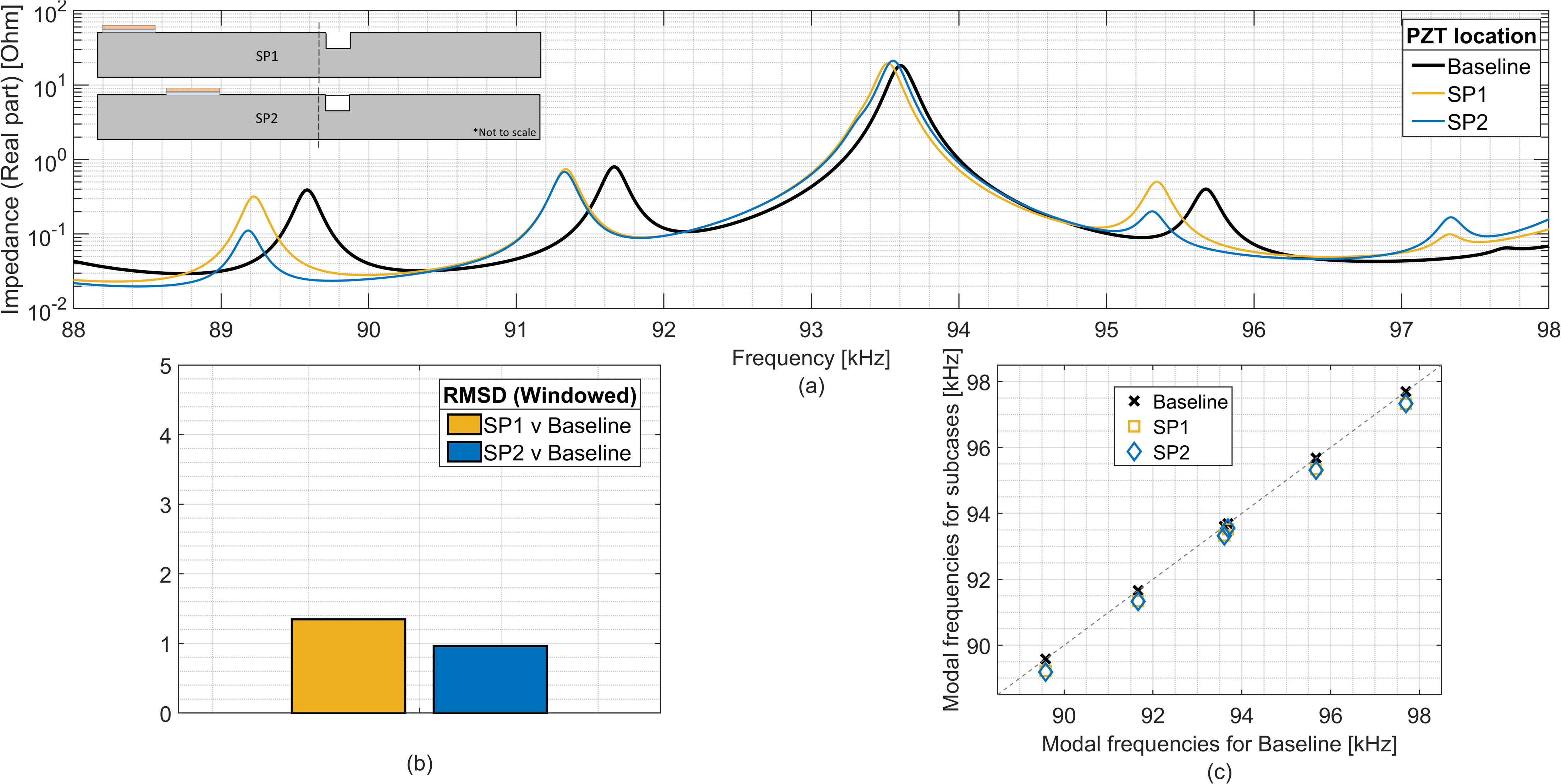}
    \caption{(a) SEM generated EMI measurements over 88 - 98 kHz, (b) Windowed-RMSD metric, and (c) VF estimated modal frequencies for Case IV. }
    \label{fig:diffpztloc_zoomed}
    \end{figure}

\textbf{Case IV} \hspace{10mm} For this example, beams with identical defects ($b$ = 5mm, $\alpha$ = 0.6) are simulated with transducers at different locations. This corresponds to a situation wherein investigations for damage on a specimen are being made using two different transducers. For SP1, the PZT is attached 60 mm from the left end while for SP2, the PZT is attached 90 mm from the left end. For both cases, the crack is located at 205mm from the left edge of the host beam. Figure \ref{fig:diffpztloc_zoomed} shows the corresponding simulated responses for a representative range of 88 - 98 kHz. 

As observed in Figure \ref{fig:diffpztloc_zoomed}, the windowed RMSD-based metric values for SP1 and SP2 are 1.35 and 0.97, respectively. Ideally, when investigations are performed using two different transducers under similar PZT-host structure interaction conditions, the measurements should remain the same. This is equivalent to performing a modal test on a specimen at two different input-output pairs. For two different input-output pairs, the recorded response might vary in amplitude for certain modal peaks, but their location should remain unchanged. Since the RMSD-based values are affected by the overall nature of the measurement, the reported values can be deceptive in indicating the presence of damage.

The estimated modal frequencies for the previously selected range of 88 - 98 kHz are plotted against those for the baseline and listed in Table \ref{tab:scenarioIV_polestablenew}, along with corresponding poles and modal damping. The modal frequencies for SP1 and SP2 are very close to one another and indistinguishable in some locations. This is in agreement with the premise of modal peak locations being insensitive to the location of the investigation. Now, when comparing the modal frequencies for either SP1 or SP2 with those for the baseline, one can infer that the particular damage causes a decrease in modal frequencies, which can be conceptually related to a reduction in stiffness value. On the other hand, even though SP1 and SP2 are investigative measurements recorded for the same damaged specimen, their reported RMSD-based metric values are indicative of a different degree of damage. Thus, unlike the RMSD-based metric, VF-estimated poles are independent of the sensor location as they are related to the sensor location agnostic modal parameters.

\begin{figure}[h!]
    \centering
    \includegraphics[width = 1\textwidth]{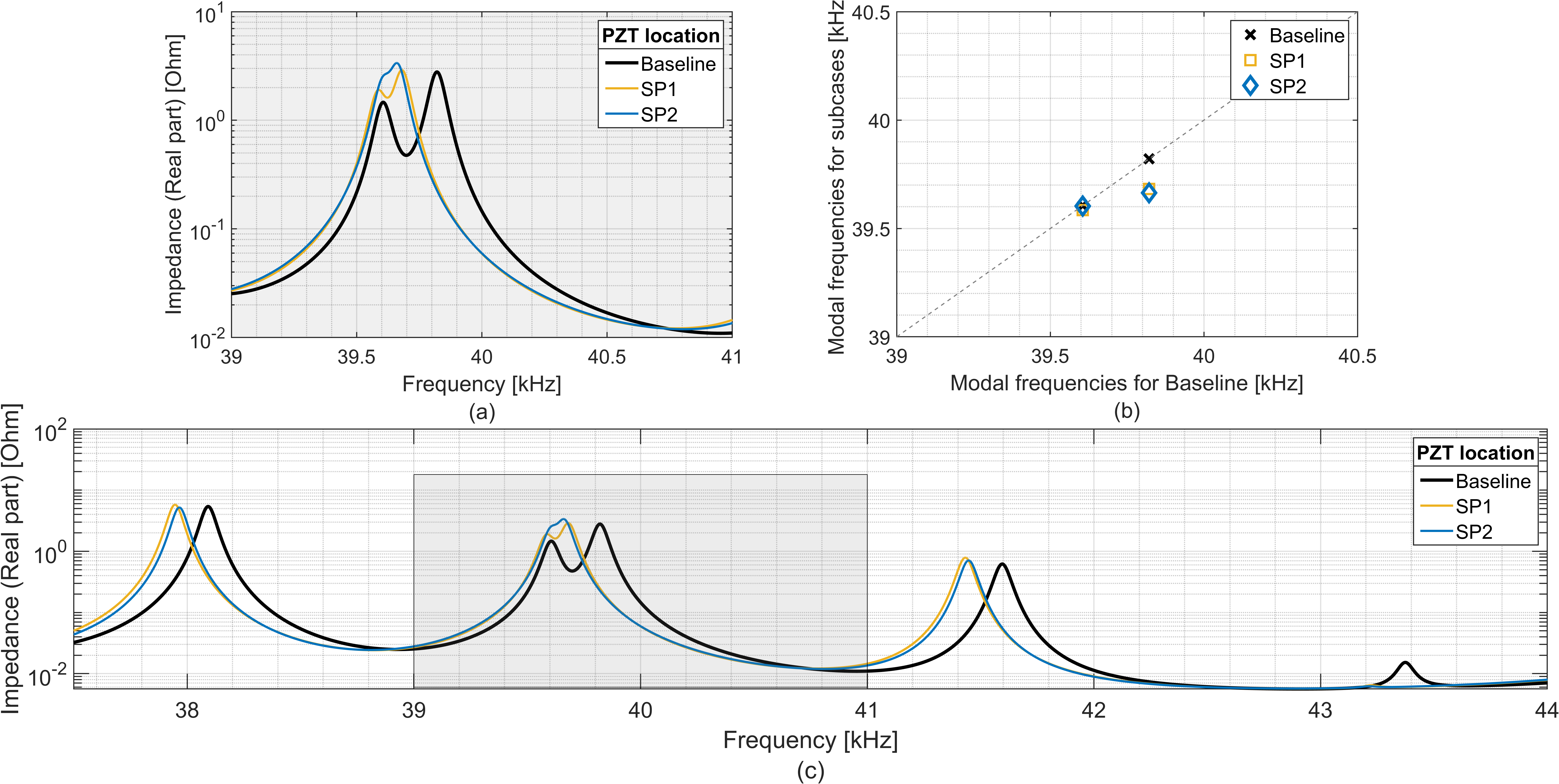}%
    \caption{(a) Modal peaks adjacent to one another, (b) VF estimated modal frequencies, and (c) SEM generated EMI measurements over 37 - 44 kHz for Case IV.}
    \label{fig:diffpztloc_VFadvantage}
\end{figure}

\newpage
Another advantage of using VF is its ability to estimate the modal peaks with high damping. Figure \ref{fig:diffpztloc_VFadvantage} shows the simulated response for SP1 and SP2 over a frequency range of 37 - 44 kHz. It can be observed that both SP1 and SP2 have two modes around 39.5 kHz, that are very close to one another. For such modes, peak picking methods \citep{ogorman_converging_1983} might miss capturing modal peaks with high damping. However, as seen in Figure \ref{fig:diffpztloc_VFadvantage}, VF is robust enough to estimate such modes despite them being close to and almost overlapping on one another.

\subsection{Experimental examples} \label{subsec:experimental}
\begin{figure}[h!]
    \centering
    \includegraphics[width = 1\textwidth]{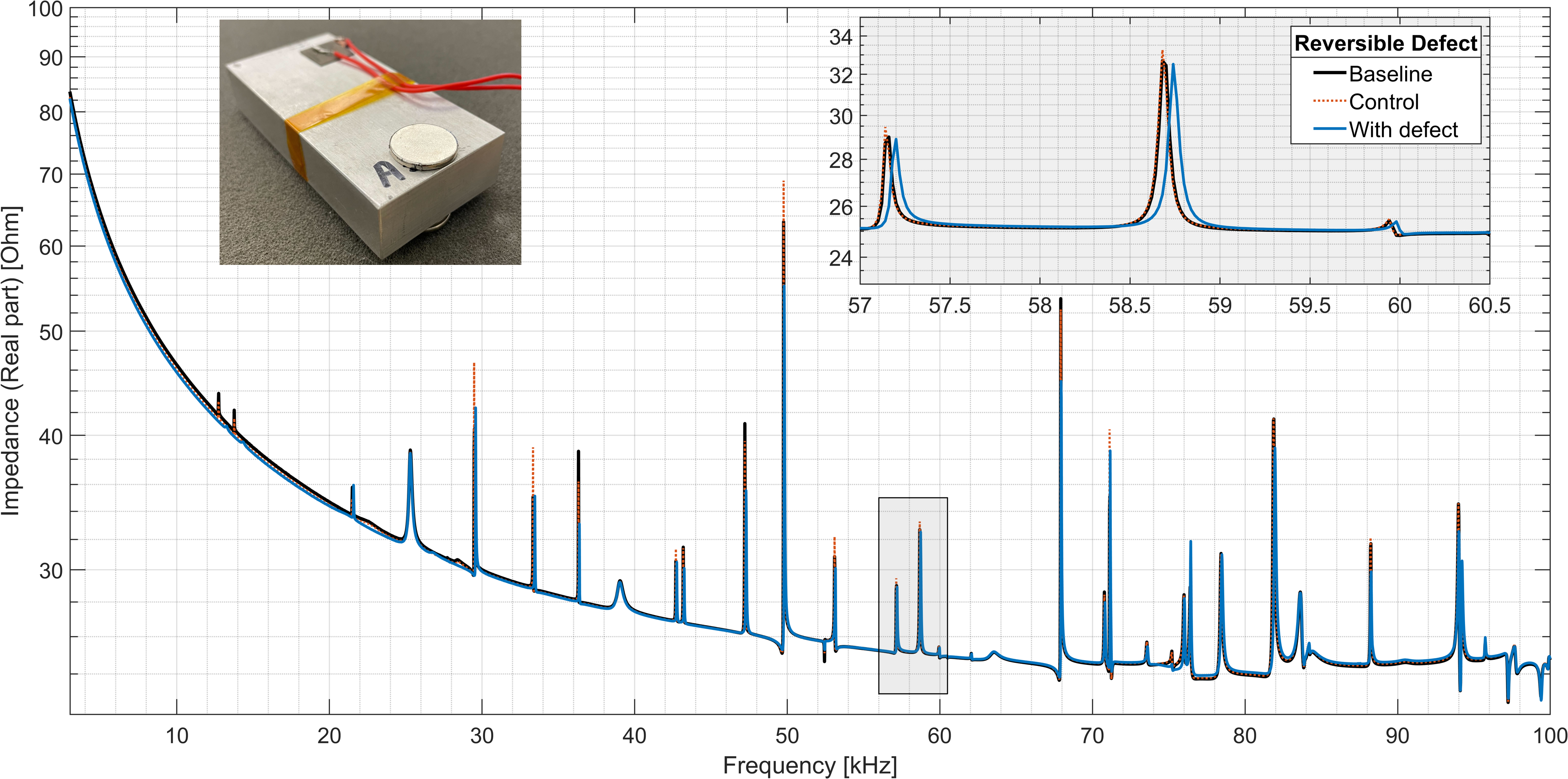}
    \caption{EMI measurements recorded via Keysight E4990A for Case V [Zoomed over 57 - 61 kHz].}
    \label{fig:beamwithmag}
\end{figure}

\textbf{Case V} \hspace{10mm} For this example, reversible damage (attaching magnets with $<5\%$ mass of specimen) was introduced into the aluminum specimen described in Section \ref{subsec:LSCF}. The magnets were attached on the edge opposite the PZT. Two EMI measurements were recorded for the baseline undamaged setting, and one was where a defect was introduced. A control measurement was defined using additional measurements for the undamaged setting. Since the previous cases were simulated, a control measurement was not essential as simulated responses are coincidental for identical parameters. However, in practice, a control measurement is a prerequisite to establishing a suitable threshold for damage detection. 
Figure \ref{fig:beamwithmag} shows a global shift in the EMI for a chosen representative range of 57 - 61 kHz. 

\begin{figure}[h!]
    \centering
    \includegraphics[width = 1\textwidth]{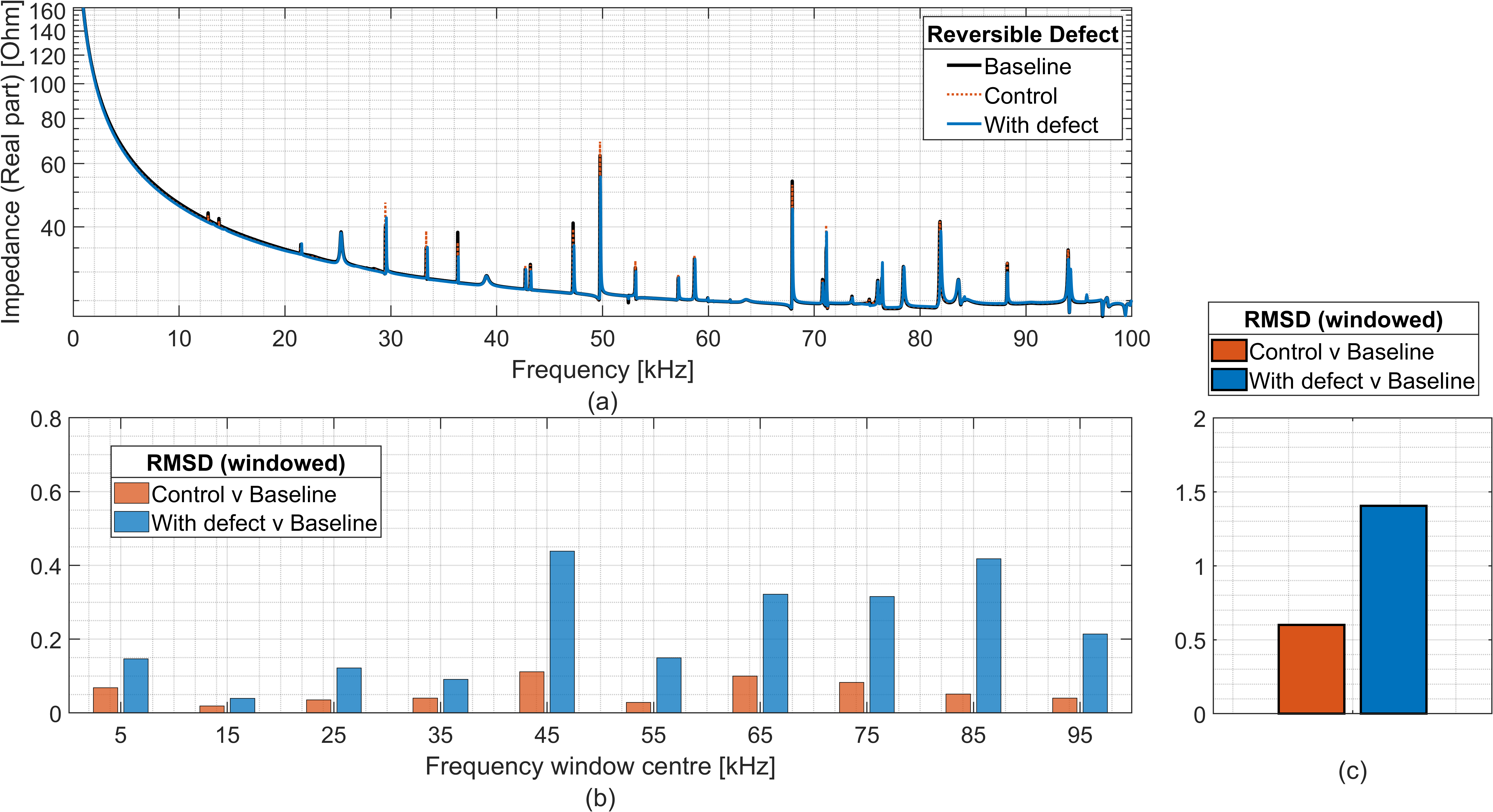}
    \caption{(a) SEM generated EMI measurements over 950 Hz - 100 kHz, (b) Standard-RMSD metric, and (c) Windowed-RMSD metric for Case V.}
    \label{fig:beamwithmag_rmsd}

\end{figure}

The recorded measurements, the corresponding standard, and windowed metric values for the recorded EMI measurements are shown in Figure \ref{fig:beamwithmag_rmsd}. The standard metric values were $0.63$ and $2.30$ for the control and the beam with defect, respectively. The windowed metric values were $0.11$ and $0.45$ for the window centered at 45 kHz while, $0.01$ and $0.35$ for the window centered around 65 kHz. Despite the unequal variation, the windowed metric values, like the standard metric values, tend to be higher for the measurement of the beam with a defect than the control measurement. As the metric values are higher than the threshold, the presence of a damage/defect can be established. Apart from the confirmation of the presence of damage, no other deduction about the physical state of the beam can be drawn just by studying the RMSD-based damage metrics. 

\begin{figure}[h!]
    \centering
    \includegraphics[width = 1\textwidth]{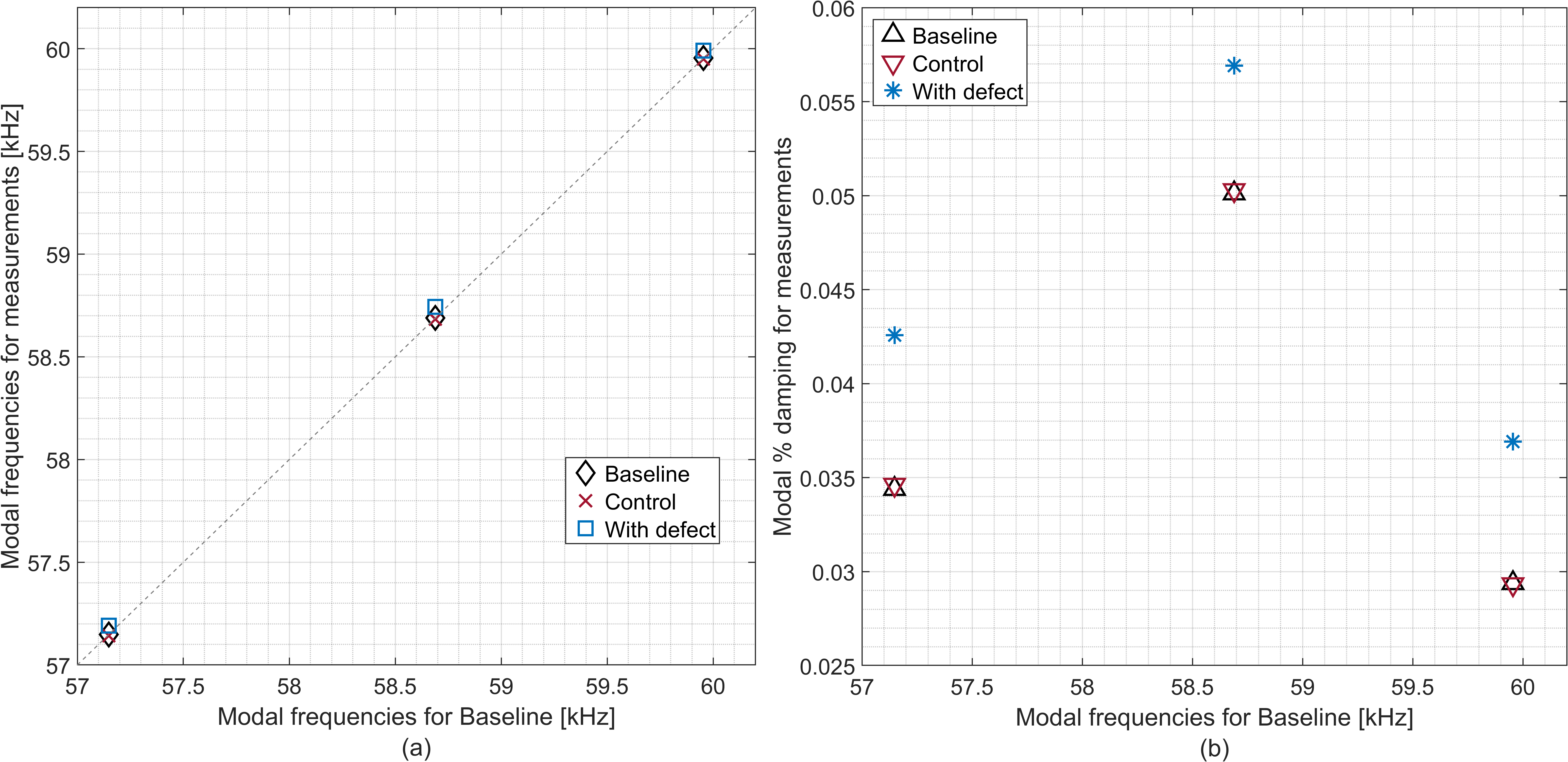}
    \caption{VF estimated (a) modal frequencies and (b) damping ratios for Example V [57 - 61 KHz].}
    \label{fig:modalfreqs_exp}
\end{figure}

Figure \ref{fig:modalfreqs_exp} shows the VF estimated frequencies and corresponding damping ratios for the beam with defect, and the control plotted along with those of the baseline over the earlier chosen range. As poles are used to approximate the frequency response function, one can estimate the frequencies as well as the damping ratios (yet another physical attribute of the system). The previous examples have focused on natural frequency estimation, but here the authors also introduce an advantage of estimating the complex pole and its extended use. Thus, along with frequencies, variations in the damping ratio can be used to make remarks about the defect(s) in the specimen under investigation. The estimated modal frequencies and the damping ratios are presented in the Appendix under Table \ref{tab:beamwithmag_polestable}. It can be observed that all estimated modal frequencies increased by about $0.1 \%$ while the estimated modal damping ratios increased on average by $10\%$ after the introduction of the defect (added mass). By preliminary inspection, one can deduce that the introduced defect caused relatively minor stiffening and considerable dampening effects. A similar minor stiffening effect can be observed in the recorded EMI measurements with added magnets as a form of indirect damage \citep{pitchford_impedance-based_2007}. One of the causes behind this can be attributed to changes in boundary conditions due to the introduced defect. A constraining boundary condition can lead to increased stiffness as well as increased damping. Further investigation can help one get a better idea of exact changes that might occur in the physical characteristics, e.g., changes in inertia, stress, etc, due to the nature of the introduced defect. 

\begin{figure}[h!]
    \centering
    \includegraphics[width = 1\textwidth]{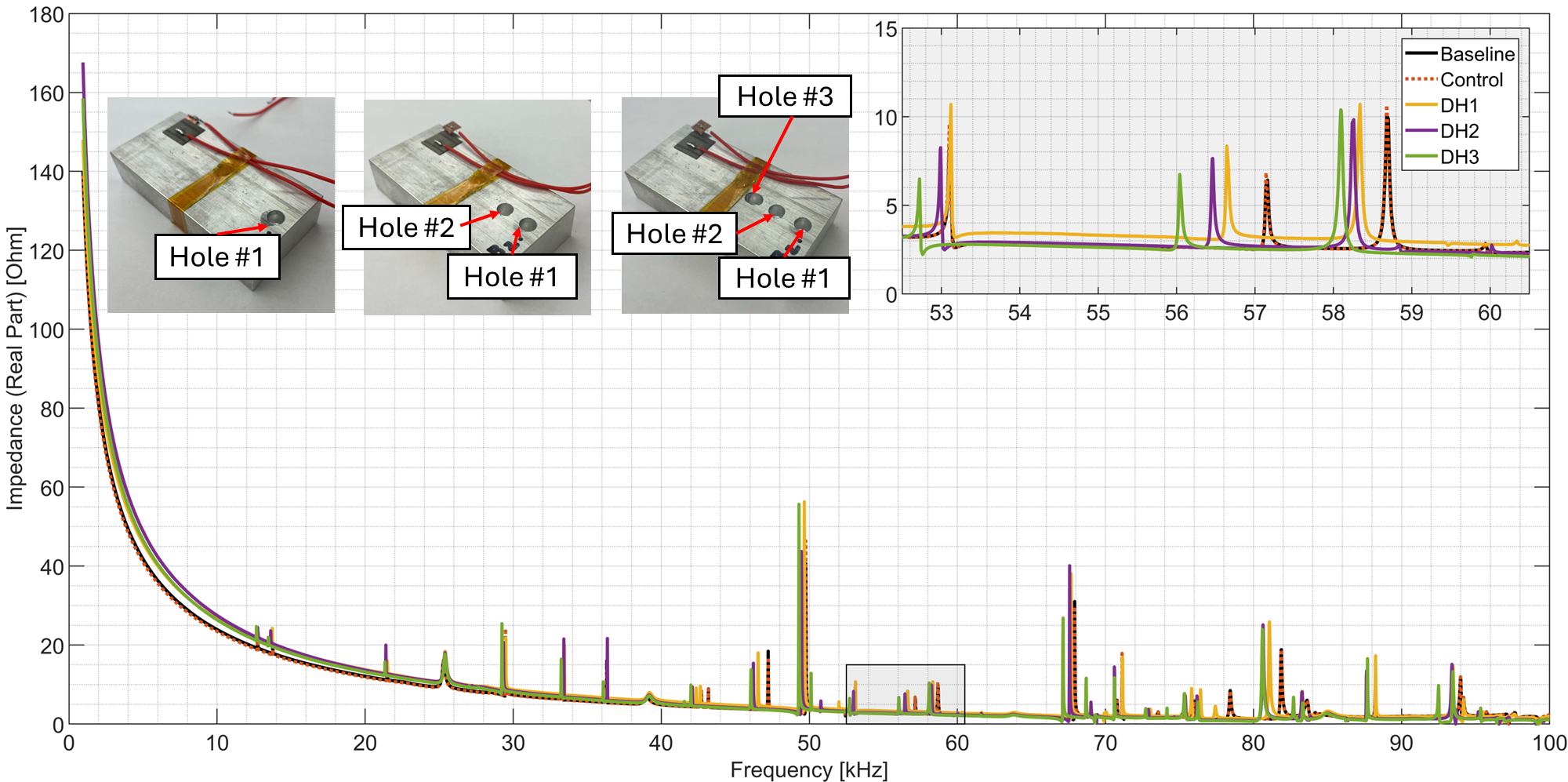}
    \caption{EMI measurements recorded via Keysight E4990A for Case VI [Zoomed over 52 - 61 kHz].}
    \label{fig:beamwithholes}
\end{figure}

\textbf{Case VI} \hspace{10mm}  For this example, progressive damage in the form of holes was introduced to an aluminum specimen. Three subsequent holes of 5.64 mm diameter each were made with the first hole's center being 6.4 mm apart from the edge opposite the PZT, and each center about 9.54 mm apart. EMI measurements were recorded after the introduction of each hole. The damage subcases with increasing holes are designated as DH1, DH2 and DH3. Examples such as these have been extensively used in the literature to showcase the potential of EMI as an SHM technique \citep{pitchford_impedance-based_2007,Yang_Liu_Annamdas_Soh_2009,singh2018quantification}. Here this example is used to showcase how VF can be used on data from a physical sample with induced progressive damage. In Figure \ref{fig:beamwithholes}, leftward shifts in peaks can be observed in the recorded EMI measurements for the representative window of 52 - 61 kHz.

Using the proposed approach, the poles, modal frequencies, and damping ratios are estimated for each subcase. For the chosen representative window, the estimated modal frequencies are shown in Figure \ref{fig:modalfreqs_exp2} and tabulated along with poles and damping ratio in the Appendix under Table \ref{tab:beamwithholes_polestable}. From DH1 to DH3, the modal frequencies experience a general decrease in values. It can be observed that all estimated modal frequencies decreased on average by $0.49 \%$, $0.74 \%$, and $1.22 \%$ after the introduction of the defect(s). Although the changes in the estimated modal damping ratios are significant, they do not follow an observable pattern. Access to mode shapes can assist in understanding the changes in the damping ratio. Thus, it can be deduced that the damage worsens from DH1 to DH3 and could potentially be interpreted as a reduction in stiffness.
\begin{figure}[h!]
    \centering
    \includegraphics[width = 1\textwidth]{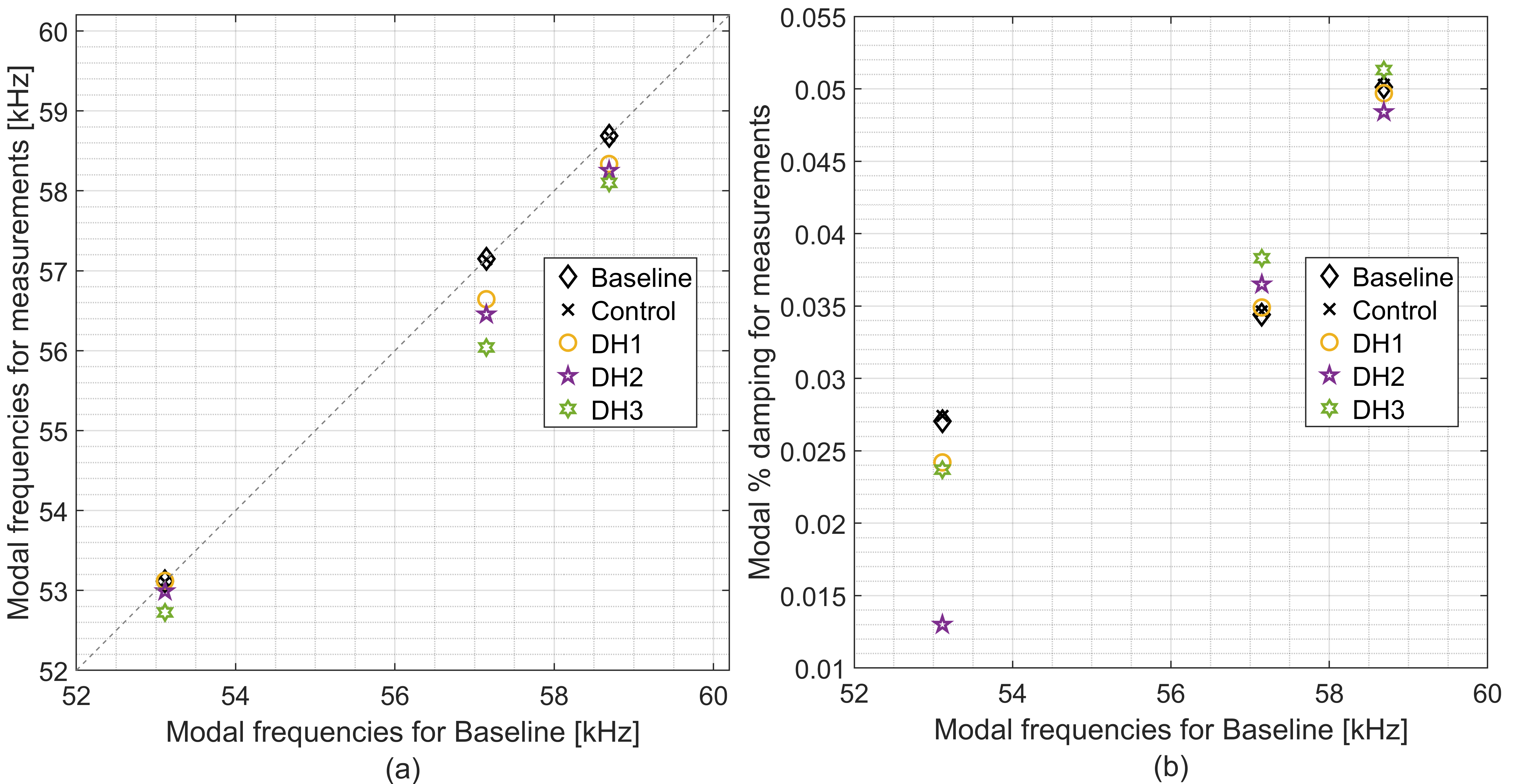}
    \caption{VF estimated (a) modal frequencies and (b) damping ratios for Case VI [52 - 61 KHz].}
    \label{fig:modalfreqs_exp2}
\end{figure}

For all the above examples, it is evident that the traditional metrics do help qualify if a specimen under investigation has undergone a change (i.e., go, no-go approach). However, when attempting to move along the stages of SHM, the traditionally used metrics alone fall short in making remarks about the physical aspects of the defect, as they were never designed with this goal in mind. Furthermore, the size and location of the chosen window affect the metric values and thus, the interpretation of the changes becomes frequency-window dependent. 
When comparing standard metric values between Cases I through III, a generalized incremental trend is observed. This opens the damage interpretation to speculations and at times to misdiagnosis. In some scenarios, e.g. Case IV, the metric value can also be deceptive and falsely qualify something as damaged. However, the VF-estimated poles, and by extension the modal parameters can help the user better understand the underlying physics of the structure. The use of these modal parameters can be implemented to move towards robust damage metrics as they tend to provide more information on the change and are agnostic to frequency range selection or other user-dependent interpretations. For Cases I through IV, the shifts in the modal frequencies were used as a metric to move a step towards understanding changes in the structure. The advantage of estimating the complex pole over the traditional metrics or even frequency estimating peak-picking methods \citep{ogorman_converging_1983}, is that it provides information about the damping ratio. The simulated subcases discussed in Cases I through IV do not showcase significant changes in the damping ratios ($\Delta < 1 \%$), and hence are not used as an additional metric. However, Case V, an experimental case study shows significant changes in the damping ratios. Thus, in scenarios similar to Case V and VI, the damping ratio can also be used along with the natural frequencies to make remarks about the defect. As the VF-estimated parameters are more robust and indicative of physical changes, the metrics based on these parameters can assist in moving to advanced stages of SHM.


\section{Conclusion} \label{sec:ConclusionFutureWorks}
The current study proposes using a pole-based approach for damage assessment in electromechanical impedance-based structural health monitoring, specifically Vector Fitting. The main conclusions of the study are as follows 

\begin{itemize}

\item Insights into understanding SHM and the limitations of the traditional damage metrics were discussed. From the literature it can be observed that traditional metrics produce scalar value by adopting an aggregate frequency approach, which can miss necessary crucial information. The metrics provide minimum intuition about damage which can be subjective to user-defined bandwidth and material of the structure.

\item A pole-based approach via VF for EMI-based structural health monitoring was introduced. VF was compared with system identification and similar mathematical reduction techniques; namely, LSCF, RKFIT, and AAA. It was shown that VF is more suitable than the rational fraction polynomial-based LSCF domain method for high-frequency EMI applications. VF was successful at estimating the EMI measurements over the entire frequency range of operation, whereas LSCF struggled even after adopting a strategy to divide the entire range into connected sections with lower modal peak density. Likewise, when compared to RKFIT and AAA, VF generated complex conjugate pole pairs that are more relevant to physical structures. Even though RKFIT and AAA successfully estimated the high-frequency EMI measurement, they produced either non-conjugate or unstable poles (right half plane) which are not representative of the physical structure.

\item The use of the proposed approach was then demonstrated through simulated and experimental examples. Comparisons of the traditional metrics with the VF-based metrics for these examples were also presented. When comparing the windowed RMSD-based metrics and VF-estimated modal parameters over smaller frequency ranges, the latter provided more qualitative information than the former (i.e., captured softening and stiffening effects, and mode-specific changes).
Information about damage effects on or around individual modal peaks can prove useful in understanding the changes in the underlying system. For a chosen bandwidth, the traditional damage metric values obtained after windowing are highly dependent on the bandwidth, while the proposed pole-based technique focuses on modal density in the chosen bandwidth which is shown to be more stable.

\item Traditional metrics condense all the available information into a single value while VF focuses on poles that carry the critical information about the system. Processing all available information into a single, undifferentiated value seems efficient, however, it can result in the inadvertent omission of crucial information. Computing the traditional metrics requires processing every element of the data set, leading to a large amount of data required to describe the EMI measurement. On the other hand, VF uses a small representative set of data, i.e. poles to perform the same function and the data acquisition can be focused around the poles of interest to reduce acquisition time further if needed. Thus, VF produces a cleaner yet qualitative representative data set for the measurement. Although pre-computation for VF is more expensive, the post-analysis and interpretation are simpler and more physically intuitive as they are linked to modal parameters.
\end{itemize}

Future studies will address how this approach can be used in tracking pole movements and associated mode shapes, which can be extended to understanding how different types of damages affect the EMI response and help progress along the stages of SHM. For instance, future augmentation to mode shape information can lead to modes that open and close a crack, and thus why they are more sensitive. This could lead to identifying the damage locations.

\section*{Acknowledgments}
This material is based upon work supported by the National Science Foundation under Award \#1931931. Any opinions, findings, and conclusions or recommendations expressed in this material are those of the author(s) and do not necessarily reflect the views of the National Science Foundation. The authors would also like to recognize the support provided by the Byron Anderson '54 fellowship and the James J. Cain ’51 fellowship at Texas A\&M University.

\bibliographystyle{unsrtnat}
\bibliography{Refs}

\newpage
\section*{Appendix}

\begin{table}[h]
    \centering
    \caption{Analytical, LSCF estimated and VF estimated for 5 DoF system (Undamaged)}
    \label{tab:AnlytLscfVf_undamaged}
    \begin{tabular}{|c|c|c|c|}
            \hline
            & \multicolumn{3}{|c|}{Undamaged}\\
            \hline
            & Analytical & LSCF Estimated & VF estimated \\      
            \hline
            \multirow{5}[0]{*}{Est. Poles ($\alpha_{n} \pm \beta_{n}i$)} 
            & $-0.011 \pm 116.51$ & $-0.011 \pm 116.48$ & $-0.011 \pm 116.51$ \\
            & $-0.040 \pm 225.08$ & $-0.040 \pm 225.01$ & $-0.040 \pm 225.08$ \\
            & $-0.080 \pm 318.31$ & $-0.080 \pm 318.21$ & $-0.080 \pm 318.31$  \\
            & $-0.119 \pm 389.85$ & $-0.119 \pm 389.73$ & $-0.119 \pm 389.85$ \\
            & $-0.148 \pm 434.82$ & $-0.148 \pm 434.68$ & $-0.148 \pm 434.82$ \\
            \hline
            \multirow{5}[0]{*}{Est. Frequency ($\omega_{n}$) [Hz]} 
            & $116.51$ & $116.48 (-0.026\%)$ & $116.51 (0.00\%)$ \\
            & $225.08$ & $225.01 (-0.031\%)$ & $225.08 (0.00\%)$ \\
            & $318.31$ & $318.21 (-0.031\%)$ & $318.31 (0.00\%)$ \\
            & $389.85$ & $389.73 (-0.031\%)$ & $389.85 (0.00\%)$ \\
            & $434.82$ & $434.68 (-0.032\%)$ & $434.82 (0.00\%)$ \\
            \hline
            \multirow{5}[0]{*}{Est. Damping ($\zeta_{n}$)} 
            & $9.15E-05$ & $9.15E-05$ & $9.15E-05$ \\
            & $1.77E-04$ & $1.77E-04$ & $1.77E-04$ \\
            & $2.50E-04$ & $2.50E-04$ & $2.50E-04$ \\
            & $3.06E-04$ & $3.06E-04$ & $3.06E-04$ \\
            & $3.42E-04$ & $3.42E-04$ & $3.42E-04$ \\
            \hline
    \end{tabular}
\end{table}

\begin{table}[h!]
    \centering
    \caption{Analytical, LSCF estimated and VF estimated for 5 DoF system (Damaged)}
    \label{tab:AnlytLscfVf_damaged}
    \begin{tabular}{|c|c|c|c|}
            \hline
            & \multicolumn{3}{|c|}{Damaged}\\
            \hline
            & Analytical & LSCF Estimated & VF estimated \\      
            \hline
            \multirow{5}[0]{*}{Est. Poles ($\alpha_{n} \pm \beta_{n}i$)} 
            & $-0.016 \pm 94.91$ & $-0.016 \pm 94.88$  & $-0.016 \pm 94.88$ \\
            & $-0.042 \pm 218.18$ & $-0.042 \pm 218.12$ & $-0.042 \pm 218.18$ \\
            & $-0.089 \pm 308.39$ & $-0.089 \pm 308.29$ & $-0.089 \pm 308.39$  \\
            & $-0.132 \pm 367.19$ & $-0.132 \pm 367.08$ & $-0.132 \pm 367.19$ \\
            & $-0.149 \pm 426.69$ &  $-0.149 \pm 426.56$ &  $-0.149 \pm 426.69$ \\
            \hline
            \multirow{5}[0]{*}{Est. Frequency ($\omega_{n}$) [Hz]} 
            & $94.91$ & $94.88 (-0.032\%)$ & $94.91 (0.00\%)$ \\
            & $218.18$ & $218.12 (-0.028\%)$ & $218.18 (0.00\%)$ \\
            & $308.39$ & $308.29 (-0.032\%)$ & $308.39 (0.00\%)$ \\
            & $367.19$ & $367.08 (-0.030\%)$ & $367.19 (0.00\%)$ \\
            & $426.69$ & $426.56 (-0.030\%)$ & $426.69 (0.00\%)$ \\
            \hline
            \multirow{5}[0]{*}{Est. Damping ($\zeta_{n}$)} 
            & $1.66E-04$ & $1.66E-04$ & $1.66E-04$ \\
            & $1.92E-04$ & $1.92E-04$ & $1.92E-04$ \\
            & $2.87E-04$ & $2.87E-04$ & $2.87E-04$ \\
            & $3.59E-04$ & $3.59E-04$ & $3.59E-04$ \\
            & $3.50E-04$ & $3.50E-04$ & $3.50E-04$ \\
            \hline
    \end{tabular}
\end{table}

\begin{table}[h!]
    \centering
    \caption{Estimated poles and modal parameters for Example I}
    \label{tab:scenarioI_polestablenew}
    \begin{adjustbox}{width=1\textwidth}
        \begin{tabular}{|c|c|c|c|c|}
            \hline
            & Baseline & SW1 & SW2 & SW3 \\
            \hline
            \multirow{9}[0]{*}{Est. Poles ($\alpha_{n} \pm \beta_{n}i$)} 
            & $-52.21 \pm 52433.75i$ & $-52.02 \pm  52249.95i$ & $-51.85 \pm  52065.05i$ & $-51.48 \pm  51693.63i$ \\
            & $-53.27 \pm 53574.33i$ & $-52.95 \pm  53217.23i$ & $-52.78 \pm  53018.85i$ & $-52.74 \pm  52953.29i$ \\
            & $-54.05 \pm 54294.67i$ & $-53.84 \pm  54088.70i$ & $-53.65 \pm  53907.93i$ & $-53.29 \pm  53535.20i$ \\
            & $-55.88 \pm 56116.21i$ & $-55.69 \pm  55923.19i$ & $-55.50 \pm  55730.28i$ & $-55.12 \pm  55353.03i$ \\
            & $-57.76 \pm 57967.90i$ & $-57.54 \pm  57751.25i$ & $-57.34 \pm  57558.60i$ & $-56.95 \pm  57163.29i$ \\
            & $-59.68 \pm 59884.07i$ & $-59.45 \pm  59662.22i$ & $-59.25 \pm  59444.32i$ & $-58.82 \pm  59024.65i$ \\
            & $-59.90 \pm 60301.86i$ & $-59.82 \pm  60250.84i$ & $-59.60 \pm  60072.82i$ & $-58.94 \pm  59413.83i$ \\
            & $-61.61 \pm 61830.08i$ & $-61.38 \pm  61598.68i$ & $-61.17 \pm  61390.18i$ & $-60.74 \pm  60948.99i$ \\
            & $-63.53 \pm 63762.64i$ & $-63.30 \pm  63527.80i$ & $-63.07 \pm  63298.15i$ & $-62.63 \pm  62860.86i$ \\
            \hline
            \multirow{9}[0]{*}{Est. Frequency ($\omega_{n}$) [Hz]} 
            & $52433.78$ & $52249.98 (-0.35\%)$ & $52065.08 (-0.70\%)$ & $51693.657 (-1.41\%)$ \\
            & $53574.36$ & $53217.26 (-0.67\%)$ & $53018.87 (-1.04\%)$ & $52953.31 (-1.16\%)$ \\
            & $54294.70$ & $54088.73 (-0.38\%)$ & $53907.96 (-0.71\%)$ & $53535.23 (-1.40\%)$ \\
            & $56116.24$ & $55923.22 (-0.34\%)$ & $55730.31 (-0.69\%)$ & $55353.05 (-1.36\%)$ \\
            & $57967.93$ & $57751.28 (-0.37\%)$ & $57558.63 (-0.71\%)$ & $57163.32 (-1.39\%)$ \\
            & $59884.10$ & $59662.25 (-0.37\%)$ & $59444.35 (-0.73\%)$ & $59024.68 (-1.44\%)$ \\
            & $60301.89$ & $60250.87 (-0.08\%)$ & $60072.85 (-0.38\%)$ & $59413.86 (-1.47\%)$ \\
            & $61830.11$ & $61598.71 (-0.37\%)$ & $61390.21 (-0.71\%)$ & $60949.02 (-1.43\%)$ \\
            & $63762.67$ & $63527.83 (-0.37\%)$ & $63298.18 (-0.73\%)$ & $62860.89 (-1.41\%)$ \\
            \hline
            \multirow{9}[0]{*}{Est. Damping ($\zeta_{n}$)} 
            & $9.96E-04$ & $9.957E-04 (-0.003\%)$ & $9.959E-04 (0.028\%)$ & $9.959E-04 (0.025\%)$ \\
            & $9.94E-04$ & $9.950E-04 (-0.057\%)$ & $9.955E-04 (0.115\%)$ & $9.959E-04 (0.152\%)$ \\
            & $9.95E-04$ & $9.955E-04 (-0.000\%)$ & $9.953E-04 (-0.017\%)$ & $9.955E-04 (0.002\%)$ \\
            & $9.96E-04$ & $9.958E-04 (-0.003\%)$ & $9.959E-04 (0.008\%)$ & $9.957E-04 (-0.014\%)$ \\
            & $9.96E-04$ & $9.963E-04 (-0.005\%)$ & $9.962E-04 (-0.021\%)$ & $9.962E-04 (-0.014\%)$ \\
            & $9.97E-04$ & $9.965E-04 (-0.001\%)$ & $9.967E-04 (0.012\%)$ & $9.965E-04 (0.003\%)$ \\
            & $9.93E-04$ & $9.928E-04 (-0.045\%)$ & $9.922E-04 (-0.106\%)$ & $9.921E-04 (-0.122\%)$ \\
            & $9.96E-04$ & $9.965E-04 (-0.002\%)$ & $9.964E-04 (-0.003\%)$ & $9.966E-04 (0.016\%)$ \\
            & $9.96E-04$ & $9.964E-04 (-0.000\%)$ & $9.965E-04 (0.008\%)$ & $9.963E-04 (-0.003\%)$ \\
            \hline
        \end{tabular}
    \end{adjustbox}
\end{table}

\begin{table}[h!]
    \centering
    \caption{Estimated poles and modal parameters for Example II}
    \label{tab:scenarioII_polestablenew}
    \begin{adjustbox}{width=1\textwidth}
        \begin{tabular}{|c|c|c|c|c|}
            \hline
            & Baseline & SL1 & SL2 & SL3 \\
            \hline
            \multirow{7}[0]{*}{Est. Poles ($\alpha_{n} \pm \beta_{n}i$)} 
            & $-71.32 \pm 71552.83i$ & $-71.05 \pm 71283.46i$ & $-71.06 \pm 71297.30i$ & $-71.06 \pm 71286.20i$ \\
            & $-71.46 \pm 72533.34i$ & $-70.89 \pm 72001.99i$ & $-71.30 \pm 72458.58i$ & $-71.37 \pm 72306.92i$ \\
            & $-73.28 \pm 73511.41i$ & $-73.03 \pm 73252.01i$ & $-73.01 \pm 73238.17i$ & $-73.02 \pm 73246.90i$ \\
            & $-75.25 \pm 75466.54i$ & $-74.96 \pm 75176.66i$ & $-74.98 \pm 75198.75i$ & $-74.97 \pm 75183.13i$ \\
            & $-77.27 \pm 77480.57i$ & $-76.99 \pm 77193.19i$ & $-76.96 \pm 77173.07i$ & $-76.97 \pm 77188.46i$ \\
            & $-78.25 \pm 79363.85i$ & $-78.26 \pm 79151.21i$ & $-77.68 \pm 78730.38i$  & $-77.66 \pm 78938.28i$ \\
            & $-79.30 \pm 79533.64i$ &                        & $-79.03 \pm 79245.10i$ & $-79.01 \pm 79230.70i$ \\
            \hline
            \multirow{8}[0]{*}{Est. Frequency ($\omega_{n}$) [Hz]} 
            & $71552.86$ & $71283.50 (-0.38\%)$ & $71297.34 (-0.36\%)$  & $71286.24 (-0.37\%)$ \\
            & $72533.38$ & $72002.02 (-0.73\%)$ & $72458.61 (-0.10\%)$  & $72306.96 (-0.31\%)$ \\
            & $73511.45$ & $73252.05 (-0.35\%)$ & $73238.21 (-0.37\%)$ & $73246.94 (-0.36\%)$ \\
            & $75466.58$ & $75176.70 (-0.38\%)$ & $75198.79 (-0.35\%)$ & $75183.17 (-0.38\%)$ \\
            & $77480.61$ & $77193.23 (-0.37\%)$ & $77173.10 (-0.40\%)$ & $77188.50 (-0.38\%)$ \\
            & $79363.89$ & $79151.24 (0.\%)$ & $78730.42 (-0.80\%)$ & $78938.32 (-0.54\%)$ \\
            & $79533.68$ & $79242.35 (0.\%)$ & $79245.14 (-0.36\%)$ & $79230.74 (-0.38\%)$ \\
            \hline
            \multirow{8}[0]{*}{Est. Damping ($\zeta_{n}$)} 
            & $9.97E-04$ & $9.97E-04 (-0.001\%)$ & $9.97E-04 (-0.001\%)$ & $9.97E-04 (0.008\%)$ \\
            & $9.85E-04$ & $9.85E-04 (-0.068\%)$ & $9.84E-04 (-0.127\%)$ & $9.87E-04 (0.185\%)$ \\
            & $9.97E-04$ & $9.97E-04 (0.00\%)$ & $9.97E-04 (-0.004\%)$ & $9.97E-04 (-0.003\%)$ \\
            & $9.97E-04$ & $9.97E-04 (-0.005\%)$ & $9.97E-04 (-0.004\%)$ & $9.97E-04 (0.003\%)$ \\
            & $9.97E-04$ & $9.97E-04 (0.007\%)$ & $9.97E-04 (0.001\%)$ & $9.97E-04 (-0.007\%)$ \\
            & $9.86E-04$ & $9.89E-04 (-0.38\%)$ & $9.87E-04 (0.073\%)$ & $9.84E-04 (-0.207\%)$ \\
            & $9.97E-04$ &                      & $9.97E-04 (0.015\%)$ & $9.97E-04 (0.012\%)$ \\
            \hline
        \end{tabular}
    \end{adjustbox}
\end{table}

\begin{table}[h!]
    \centering
    \caption{Estimated poles and modal parameters for Example III}
    \label{tab:scenarioIII_polestablenew}
    \begin{adjustbox}{width=1\textwidth}
        \begin{tabular}{|c|c|c|c|c|}
            \hline
            &Baseline & SF1 & SF2 & SF3 \\
            \hline
            \multirow{6}[0]{*}{Est. Poles ($\alpha_{n} \pm \beta_{n}i$)} 
            & $-43.14 \pm 43373.39i$ & $-43.06 \pm 43287.21i$ & $-43.02 \pm 43239.93i$ & $-42.99 \pm 43212.13i$ \\
            & $-44.89 \pm 45117.22i$ & $-44.80 \pm 45037.45i$ & $-44.75 \pm 44991.47i$ & $-44.73 \pm 44963.72i$ \\
            & $-46.32 \pm 46538.11i$ & $-46.23 \pm 46454.92i$ & $-46.17 \pm 46403.87i$ & $-46.13 \pm 46372.06i$ \\
            & $-46.65 \pm 46860.67i$ & $-46.55 \pm 46765.46i$ & $-46.50 \pm 46714.67i$ & $-46.47 \pm 46685.28i$ \\
            & $-48.47 \pm 48662.59i$ & $-48.38 \pm 48576.36i$ & $-48.33 \pm 48524.89i$ & $-48.3o \pm 48493.30i$ \\
            & $-50.34 \pm 50538.70i$ & $-50.23 \pm 50425.86i$ & $-50.17 \pm 50367.56i$ & $-50.14 \pm 50334.44i$ \\
            \hline
            \multirow{8}[0]{*}{Est. Frequency ($\omega_{n}$) [Hz]} 
            & $43373.41$ & $43287.23 (-0.02\%)$ & $43239.95 (-0.31\%)$ & $43212.15 (-0.37\%)$ \\
            & $45117.24$ & $45037.47 (-0.18\%)$ & $44991.49 (-0.28\%)$ & $44963.74 (-0.34\%)$ \\
            & $46538.13$ & $46454.94 (-0.18\%)$ & $46403.90 (-0.29\%)$ & $46372.08 (-0.36\%)$ \\
            & $46860.69$ & $46765.48 (-0.20\%)$ & $46714.70 (-0.31\%)$ & $46685.30(-0.37\%)$ \\
            & $48662.62$ & $48576.38 (-0.18\%)$ & $48524.92 (-0.28\%)$ & $48493.32(-0.35\%)$ \\
            & $50538.72$ & $50425.88 (-0.22\%)$ & $50367.58 (-0.34\%)$ & $50334.46(-0.40\%)$ \\
            \hline
            \multirow{6}[0]{*}{Est. Damping ($\zeta_{n}$)} 
            & $9.95E-04$ & $9.95E-04 (0.019\%)$ & $9.95E-04 (0.025\%)$ & $-9.95E-04 (0.027\%)$ \\
            & $9.95E-04$ & $9.95E-04 (-0.016\%)$ & $9.95E-04 (-0.021\%)$ & $-9.95E-04 (-0.023\%)$ \\
            & $9.95E-04$ & $9.95E-04 (-0.020\%)$ & $9.95E-04 (-0.032\%)$ & $-9.95E-04 (-0.040\%)$ \\
            & $9.95E-04$ & $9.95E-04 (0.004\%)$ & $9.95E-04 (0.004\%)$ & $-9.95E-04 (0.002\%)$ \\
            & $9.96E-04$ & $9.96E-04 (-0.008\%)$ & $9.96E-04 (-0.010\%)$ & $-9.96E-04 (-0.010\%)$ \\
            & $9.96E-04$ & $9.96E-04 (0.007\%)$ & $9.96E-04 (0.009\%)$ & $-9.96E-04 (0.009\%)$ \\
            \hline
        \end{tabular}
    \end{adjustbox}
\end{table}

\begin{table}[h!]
    \centering
    \caption{Estimated poles and modal parameters for Example IV}
    \label{tab:scenarioIV_polestablenew}
    \begin{tabular}{|c|c|c|c|}
            \hline
            & Baseline & SP1 & SP2 \\
            \hline
            \multirow{5}[0]{*}{Est. Poles ($\alpha_{n} \pm \beta_{n}i$)} 
            & $-89.37 \pm 89582.13i$ & $-89.02 \pm  89221.87i$ & $-88.99 \pm  89183.17i$ \\
            & $-91.43 \pm 91663.37i$ & $-91.10 \pm  91335.19i$ & $-91.10 \pm  91326.88i$ \\
            & $-92.68 \pm 93603.62i$ & $-93.09 \pm  93330.54i$ & $-93.08 \pm  93322.79i$ \\
            & $-93.43 \pm 93686.91i$ & $-92.50 \pm  93519.47i$ & $-92.56 \pm  93554.20i$ \\
            & $-95.46 \pm 95674.42i$ & $-95.11 \pm  95342.08i$ & $-95.10 \pm  95311.99i$ \\

            \hline
            \multirow{5}[0]{*}{Est. Frequency ($\omega_{n}$) [Hz]} 
            & $89582.18$ & $89221.92 (-0.4\%)$ & $89183.22 (-0.4\%)$ \\
            & $91663.41$ & $91335.23 (-0.4\%)$ & $91326.93 (-0.4\%)$ \\
            & $93603.66$ & $93330.57 (-0.3\%)$ & $93322.83 (-0.3\%)$ \\
            & $93686.96$ & $93519.52 (-0.2\%)$ & $93554.25 (-0.2\%)$ \\
            & $95674.47$ & $95342.13 (-0.4\%)$ & $95312.04 (-0.4\%)$ \\

            \hline
            \multirow{5}[0]{*}{Est. Damping ($\zeta_{n}$)} 
            & $9.98E-04$ & $9.98E-04 (0.005\%)$ & $9.98E-04 (0.012\%)$ \\
            & $9.97E-04$ & $9.97E-04 (0.003\%)$ & $9.97E-04 (0.007\%)$ \\
            & $9.90E-04$ & $9.97E-04 (0.735\%)$ & $9.97E-04 (0.735\%)$ \\
            & $9.97E-04$ & $9.89E-04 (-0.822\%)$ & $9.89E-04 (-0.827\%)$ \\
            & $9.98E-04$ & $9.98E-04 (-0.008\%)$ & $9.98E-04 (0.014\%)$ \\
            \hline
    \end{tabular}
\end{table}

\begin{table}[h!]
    \centering
    \caption{Estimated poles and modal parameters for Example V}
    \label{tab:beamwithmag_polestable}
    \begin{tabular}{|c|c|c|c|}
            \hline
            & Baseline & Control & With defect\\
            \hline
            \multirow{3}[0]{*}{Est. Poles ($\alpha_{n} \pm \beta_{n}i$)} 
            & $-19.66 \pm 57147.79i$ & $-19.79 \pm 57142.68i$ & $-24.35 \pm 57191.10i$ \\
            & $-29.41 \pm 58689.25i$ & $-29.51 \pm 58683.87.79i$ & $-33.44 \pm 58743.29i$ \\
            & $-17.62 \pm 59954.65i$ & $-17.58 \pm 59949.48i$ & $-22.15 \pm 59991.04i$ \\
            \hline
            \multirow{3}[0]{*}{Est. Frequency ($\omega_{n}$) [Hz]} 
            & $57147.79$ & $57142.68 (-0.01\%)$ &$57191.10 (0.08\%)$ \\
            & $58689.25$ & $58683.88 (-0.01\%)$ &$58743.30 (0.09\%)$ \\
            & $59954.65$ & $59949.48 (-0.01\%)$ &$59991.04 (0.06\%)$  \\
            \hline
            \multirow{3}[0]{*}{Est. Damping ($\zeta_{n}$)} 
            & $3.441E-04$ & $3.462E-04 (0.64\%)$ &$4.258E-04 (23.76\%)$  \\
            & $5.012E-04$ & $5.029E-04 (0.35\%)$ &$5.692E-04 (13.58\%)$ \\
            & $2.939E-04$ & $2.933E-04 (-0.23\%)$ &$3.692E-04 (25.60\%)$ \\
            \hline
    \end{tabular}
\end{table}

\begin{table}[h!]
\scriptsize	
    \centering
    \caption{Estimated poles and modal parameters for Case VI}
    \label{tab:beamwithholes_polestable}
    \begin{tabular}{|c|c|c|c|c|c|}
            \hline
            & Baseline & Control & DH1 & DH2 & DH3 \\
            \hline
            \multirow{3}[0]{*}{Est. Poles ($\alpha_{n} \pm \beta_{n}i$)} 
            & $-14.37 \pm 53113.40i$ & $-14.56 \pm 53107.83i$ & $-12.85 \pm 53119.94i$ & $-6.87 \pm 52990.91i$ & $-12.49 \pm 52724.79i$\\
            & $-19.66 \pm 57147.79i$ & $-19.79 \pm 57142.68i$ & $-19.78 \pm 56643.64i$ & $-20.62 \pm 56454.86i$ & $-21.44 \pm 56040.12i$\\
            & $-29.41 \pm 58689.25i$ & $-29.51 \pm 58683.87i$ & $-29.01 \pm 58335.40i$ & $-28.17 \pm 58251.61i$ & $-29.79 \pm 58099.51i$ \\
            \hline
            \multirow{3}[0]{*}{Est. Frequency ($\omega_{n}$) [Hz]} 
            & $53113.40$ & $53107.83 (-0.01\%)$ & $53119.94 (0.01\%)$ & $52990.92 (-0.23\%)$ & $52724.79 (-0.73\%)$ \\
            & $57147.79$ & $57142.68 (-0.01\%)$ & $56643.65 (-0.88\%)$ & $56454.86 (-1.21\%)$ & $56040.13 (-1.94\%)$ \\
            & $58689.25$ & $58683.88 (-0.01\%)$ & $58335.40 (-0.60\%)$ & $58251.61 (-0.75\%)$ & $58099.52 (-1.00\%)$\\
            \hline
            \multirow{3}[0]{*}{Est. Damping ($\zeta_{n}$)} 
            & $2.705E-04$ & $2.711E-04 (0.20\%)$ & $2.42E-04 (-10.6\%)$ & $1.30E-04 (-10.6\%)$ & $2.37E-04 (-12.4\%)$  \\
            & $3.441E-04$ & $3.462E-04 (0.64\%)$ & $3.49E-04 (1.5\%)$ & $3.65E-04 (-10.6\%)$ & $3.83E-04 (11.2\%)$  \\
            & $5.012E-04$ & $5.029E-04 (0.35\%)$ & $4.84E-04 (-0.8\%)$ & $4.84E-04 (-10.6\%)$ & $5.13E-04 (2.3\%)$\\
            \hline
    \end{tabular}
\end{table}

\end{document}